\documentclass{svjour3}                     % onecolumn (standard format)
\smartqed  % flush right qed marks, e.g. at end of proof
\usepackage{graphicx}
\usepackage{multirow}
\usepackage{makecell}
\usepackage[authoryear]{natbib}
\usepackage{url}
\usepackage[utf8]{inputenc}
\usepackage{amssymb}
\usepackage{array}
\usepackage{kantlipsum}
\usepackage{subcaption}
\usepackage{tabularx}
\usepackage{color}
\usepackage{xcolor}
\usepackage[switch]{lineno}
\usepackage{fix-cm}
\usepackage{amsmath}
\usepackage{lscape}
\newcommand{\ns}{\emph{$\nu$Solve}}
\begin{document}
%%%%%%%%%%%%%%%%%%%%%%%%%%%%%%%%%%%%%%%%%%%%%%%%%%%%%%%%%%%%%%%%%%%%%%%%%%%%%
\title{Short-baseline interferometry local-tie experiments at the Onsala Space Observatory}

%\thanks{}
% Grants or other notes about the article that should go on the front
% page should be placed within the \thanks{} command in the title
% (and the %-sign in front of \thanks{} should be deleted)
%
% General acknowledgments should be placed at the end of the article.

\subtitle{}

\titlerunning{Interferometric local-ties at OSO}        % if too long for running head

\author{Eskil Varenius \and R{\"u}diger~Haas \and Tobias Nilsson}

%\authorrunning{Short form of author list} % if too long for running head

\institute{E.~Varenius, R.~Haas 
\at
Chalmers University of Technology \\
Department of Space, Earth and Environment\\
Onsala Space Observatory \\
SE-439~92 Onsala\\
Tel.: +46-31-772-5575\\
Fax: +46-31-772-5590\\
\email{eskil.varenius@chalmers.se, rudiger.haas@chalmers.se}\\
\\
T.~Nilsson\\
Lantm{\"a}teriet -- The Swedish Mapping, Cadastral, and Land Registration Authority\\
Lantm{\"a}terigatan 2C\\
SE-801~82 G{\"a}vle, Sweden\\
}

\date{Received: date / Accepted: date}
% The correct dates will be entered by the editor

\maketitle

%- - - - - - - - - - - - - - - - - - - - - - - - - - - - - - - - -
\begin{abstract}
%- - - - - - - - - - - - - - - - - - - - - - - - - - - - - - - - -
We present results from observation, correlation and  analysis of interferometric measurements between the three geodetic very long baseline interferometry (VLBI) stations at the Onsala Space Observatory.
In total 25 sessions were observed in 2019 and 2020, most of them 24~hours long, all using X~band only.
These involved the legacy VLBI station ONSALA60 and the Onsala twin telescopes, ONSA13NE and ONSA13SW, two broadband stations for the next generation geodetic VLBI global observing system (VGOS). 
We used two analysis packages: \ns{} to pre-process the data and solve ambiguities, and \emph{ASCOT} to solve for station positions, including modelling gravitational deformation of the radio telescopes and other significant effects.
We obtained weighted root mean square postfit residuals for each session on the order of 10--15~ps using group delays and 2--5~ps using phase delays. 
The best performance was achieved on the (rather short) baseline between the VGOS stations.
As the main result of this work we determined the coordinates of the Onsala twin telescopes in VTRF2020b with sub-millimeter precision.
This new set of coordinates should be used from now on for scheduling,  correlation, as a~priori for data analyses, and for comparison with classical local-tie techniques.
Finally, we find that positions estimated from phase-delays are offset $\sim+3$~mm in the Up-component with respect to group-delays. Additional modelling of (elevation-dependent) effects may contribute to future understanding of this offset.

%- - - - - - - - - - - - - - - - - - - - - - - - - - - - - - - - -
\end{abstract}
%- - - - - - - - - - - - - - - - - - - - - - - - - - - - - - - - -

%- - - - - - - - - - - - - - - - - - - - - - - - - - - - - - - - -
\section{Introduction}
\label{Introduction}
%- - - - - - - - - - - - - - - - - - - - - - - - - - - - - - - - 
Geodetic very long baseline interferometry (VLBI) is a space geodetic technique that is of major importance \citep{Bachmann2016} for the International Terrestrial Reference Frame (ITRF) \citep{AltamimiITRF2014}.
The ITRF is the most precise and accurate realization of a global geodetic reference frame (GGRF) that is needed by the scientific community as well as society as large.
Its importance for a sustainable development of human society has been highlighted in a corresponding United Nation resolution \citep{UN2017}.

The basis for the creation and maintenance of the ITRF are space geodetic observations that are performed at so-called geodetic core (fundamental) sites.
Geodetic core sites are equipped with co-located instrumentation for a variety of space geodetic measurements, such as e.g. geodetic VLBI and Global Navigation Satellite Systems (GNSS).
To make the best possible use of the different space geodetic measurements at co-location sites, the local geometrical relations between the reference points of the different space geodetic  instrumentation has to be known with high accuracy.

The Onsala Space Observatory (OSO) is one of these geodetic core sites and operates several co-located space geodetic instruments.
Since 1979 the 20~m diameter radio telescope, ONSALA60 (ON), is used for geodetic VLBI and contributes regularly to the observing program of the International VLBI Service for Geodesy and Astrometry (IVS).
This makes ONSALA60 the station with the longest time series of observations of all the VLBI stations that are active in the IVS.% \textbf{(REF?)}.

During recent years, two new 13.2~m diameter radio telescopes were built at Onsala Space Observatory, ONSA13NE (OE) and ONSA13SW (OW), the so-called Onsala twin telescopes (OTT)
\citep{Haas_2013, Haas_et_al_2019}.
The OTT are instruments for the next generation VLBI system \citep{Petrachenko_et_al_2009}, commonly referred to as VGOS (VLBI Global Observing System). 
VGOS is going to become the work horse of the IVS for the coming decades.
Due to very fast slewing telescopes with insignificant structural deformation and dual-polarized broadband receivers, VGOS is expected to improve the performance by one order of magnitude compared to the so-called legacy S/X VLBI system \citep{Petrachenko_et_al_2009}, and thus be able to reach the accuracy level that is necessary to address the societal needs in connection with the Global Geodetic Observing System \citep[GGOS;][]{Plag2009} and global change research in general.

In order to use the new VGOS telescopes and observations together with the other existing space geodetic observations, the so-called local tie vectors between the reference points of the new and the previously existing instrument have to be determined.
Usually, local tie vectors between reference points for different space geodetic instrumentation are determined with classical geodetic survey, see e.g. \citet{Haas_Eschelbach_2005, Loesler_et_al_2013, Loesler_et_al_2016}.
However, in the case of co-located instrumentation of the same space geodetic technique, also technique-inherent observations can be used to determine the local tie vectors.
For this purpose,  we performed during 2019 and 2020 a series of short-baseline local interferometry campaigns with the three co-located geodetic VLBI stations at Onsala, i.e. ON, OE and OW.

Section~\ref{SEC:systems} presents briefly the three main instruments at Onsala that are used for geodetic VLBI. 
The design and setup of the short-baseline co-location experiments performed in 2019 and 2020 are described in Section~\ref{SEC:experiments}.
Section~\ref{SEC:correlation} explains data correlation and post-correlation analysis.
The methods of geodetic analysis of the resulting delay data is presented in Section~\ref{SEC:analysis}, and the resulting positions are given in Section~\ref{SEC:results}.
Finally, Section~\ref{SEC:summary} gives a summary and outlook.

%- - - - - - - - - - - - - - - - - - - - - - - - - - - - - - - - -
\section{The geodetic VLBI systems at Onsala}
\label{SEC:systems}
%- - - - - - - - - - - - - - - - - - - - - - - - - - - - - - - - -

\begin{figure*}[htb]
\centering
\includegraphics[width=1\textwidth]{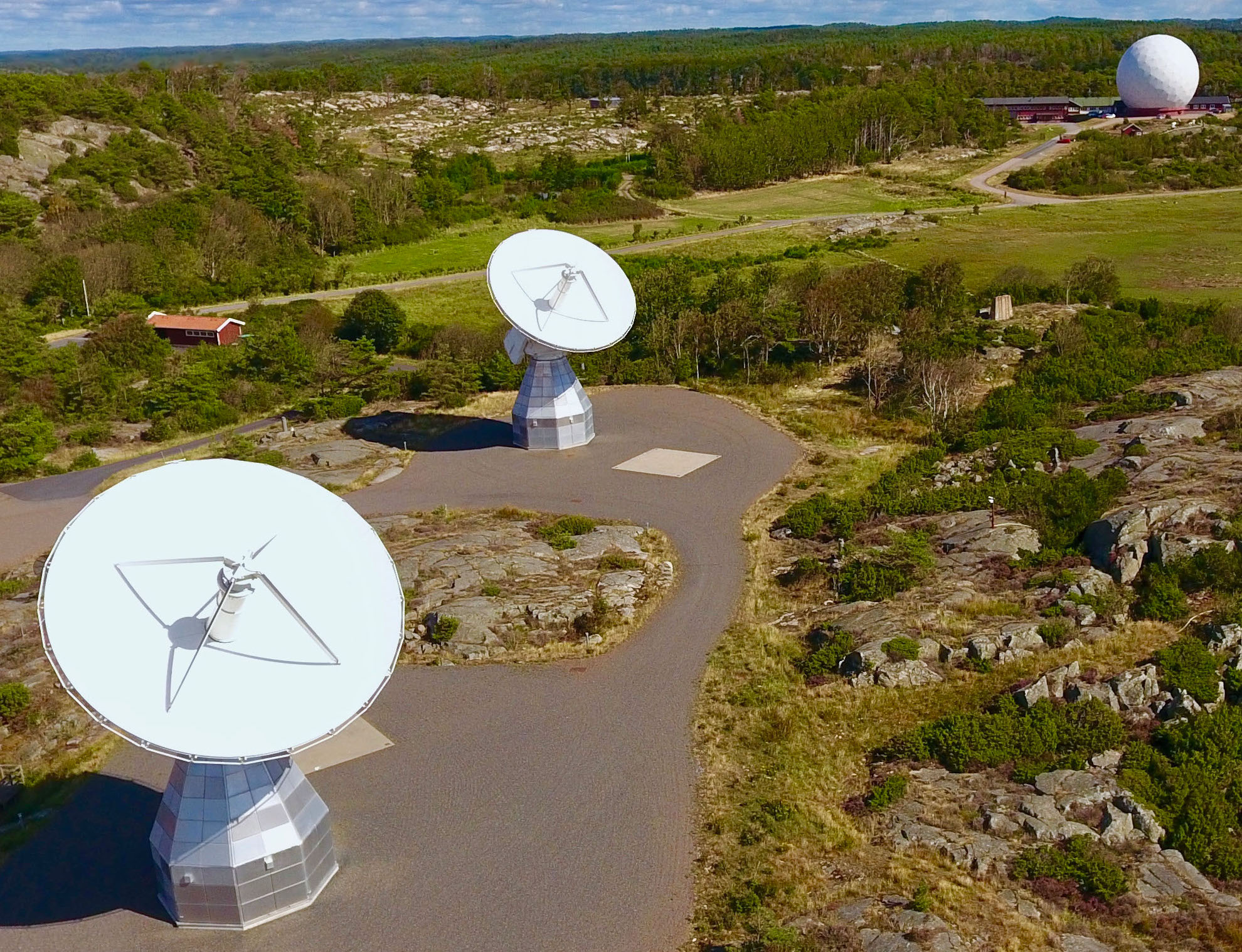}
\caption{The Onsala telescope cluster used in these observations: OW (left), OE (middle), and ON (right; inside radome). 
OW and OE are about 75~m apart, and the baseline between ON and OE is about 470~m long. 
Credit: Onsala Space Observatory/Roger Hammargren.}
\label{FIG01:OSO}      
\end{figure*}

The Onsala Space Observatory has the longest time series of VLBI observations in Europe, going back to the first astro/geodetic session already in 1968 \citep{Whitney_1974}.
In the early days, the 25~m radio telescope built in 1963, was used for VLBI.
During 1976 to 1979 the radome-enclosed 20~m radio telescope (ONSALA60, ON) was built (right in Fig.~\ref{FIG01:OSO}) and has been used since then for geodetic VLBI \citep{Scherneck_et_al_1998}.
This makes ON the VLBI station with the longest time series in the IVS.
During the last decades, ON has participated in on the order of 50 S/X sessions per year, and has been involved in all continuous (CONT) campaigns that the IVS organised.

During 2015--2017 the OTT (left and middle in Fig.~\ref{FIG01:OSO}) were installed \citep{Haas_2013, Haas_et_al_2019}, following the design for the VGOS, the next generation VLBI system \citep{Petrachenko_et_al_2009}.
The OTT were inaugurated in May 2017 and in late 2017 they started to participate in first international VGOS test sessions.
After thorough system tests, the OTT are now operating regularly in the IVS VGOS sessions since early 2019.
Table~\ref{TAB:01} provides some technical parameters for comparing the three geodetic VLBI systems at Onsala.

%- - - - - - - - - - - - - - - - - - - - - - - - - - - - - - - - -
% FIG-01
%- - - - - - - - - - - - - - - - - - - - - - - - - - - - - - - - -
%- - - - - - - - - - - - - - - - - - - - - - - - - - - - - - - - -

% - - - - - - - - - - - - - - - - - - - - - - - - - - - - 
% TAB-1
% - - - - - - - - - - - - - - - - - - - - - - - - - - - - 
\begin{table}[htb]
\centering
\caption{The VLBI stations at the Onsala Space Observatory used for geodetic VLBI observations.}
\begin{tabular}{ l | l | l | l | l}
\hline
VLBI station  & ID & diameter & geodetic frequency range & polarization\\ 
\hline
ONSALA60 & ON & 20.0~m & 2.2 -- 2.4~GHz and 8.1 -- 9.0~GHz  & R / R + L\\
ONSA13NE & OE & 13.2~m & 3.0 -- 15~GHz  & X + Y\\
ONSA13SW & OW & 13.2~m & 2.2 -- 14~GHz  & X + Y \\
\hline
\end{tabular}
\label{TAB:01}
\end{table}
% - - - - - - - - - - - - - - - - - - - - - - - - - - - - 

%- - - - - - - - - - - - - - - - - - - - - - - - - - - - - - - - -
\section{ONTIE observations}
\label{SEC:experiments}
%- - - - - - - - - - - - - - - - - - - - - - - - - - - - - - - - -

We started to gain first experience with short-baseline experiments at Onsala in early 2019 as part of a master's thesis project \citep{Marknas_2019}.
The longest distance between the telescopes is only about 550~m which is short enough for differential ionospheric effects to be negligible, and therefore there is no need for dual-band (e.g. S/X) observations.
Since the common overlapping frequency range for all three systems is 8.1--9~GHz (X-band), we focused entirely on this band. 
In this work we analyse the 25 VGOS databases (vgosDb) summarised in Table~\ref{TAB:obs}.

% - - - - - - - - - - - - - - - - - - - - - - - - - - - - 

% - - - - - - - - - - - - - - - - - - - - - - - - - - - - 
% TAB-2
% - - - - - - - - - - - - - - - - - - - - - - - - - - - - 
\begin{table}[htb]
\centering
\caption{Summary of the 25 vgosDBs analysed in this work. 
Some experiments spanned several days, and were split in multiple vgosDbs of approximately 24~h. Column 3 mark channels excluded from fringe-fitting due to broadband RFI; channels in parenthesis only excluded on the OE-OW baseline.}
\begin{tabular}{ l | c | l | l | r |l |l }
\hline
Exp. & Conf. in& Excl. & vgosDB & Dur. & Stations & Manual \\ 
code & Table~\ref{TAB:freqs} & chans.& & [h]&  & PCAL  \\
\hline
ON9114 & A & -     & 19APR24VB & 4 & ON, OE & ON \\
ON9120 & A & a (b) & 19APR30VB & 29 & ON, OE, OW & ON, OW \\
ON9122 & A & a (b) & 19MAY02VB & 29 & ON, OE & ON \\
ON9135 & B & a (b) & 19MAY15VB &  4 & ON, OE, OW & ON, OW \\
ON9136 & B & a (b) & 19MAY16VB & 26 & ON, OE, OW & ON, OW\\
ON9142 & B & a (b) & 19MAY22VB &  6 & ON, OE, OW & ON, OW\\
ON9323 & B & ab    & 19NOV19VB & 21 & ON, OE, OW & -\\
ON9327 & B & ab    & 19NOV23VB & 24 & ON, OE ,OW & -\\
ON9328 & B & ab    & 19NOV24VB & 24 & ON, OE, OW & -\\
ON0010 & C & ab (c)& 20JAN10VB & 20 & ON, OE, OW & -\\
ON0011 & C & ab (c)& 20JAN11VB & 20 & ON, OE, OW & -\\
ON0012 & C & ab (c)& 20JAN12VB & 20 & ON, OE, OW & -\\
ON0079 & C & a (e) & 20MAR19VB & 24 & ON, OE, OW & -\\
ON0080 & C & a (e) & 20MAR20VB & 24 & ON, OE, OW & -\\
ON0081 & C & a (e) & 20MAR21VB & 24 & ON, OE, OW & -\\
ON0082 & C & a (e) & 20MAR22VB & 24 & ON, OE, OW & -\\
ON0177 & C & a (e) & 20JUN25VB & 23 & ON, OE, OW & -\\
ON0178 & C & a (e) & 20JUN26VB & 23 & ON, OE, OW & -\\
ON0179 & C & a (e) & 20JUN27VB & 23 & ON, OE, OW & -\\
ON0180 & C & a (e) & 20JUN28VB & 23 & ON, OE, OW & -\\
ON0223 & C & a (e) & 20AUG10VB & 24 & ON, OE, OW & ON\\
ON0227 & C & a (e) & 20AUG14VB & 24 & ON, OE, OW & -\\
ON0228 & C & a (e) & 20AUG15VB & 24 & ON, OE, OW & -\\
ON0317 & C & a (e) & 20NOV12VB & 24 & ON, OE, OW & -\\
ON0318 & C & a (e) & 20NOV13VB & 23 & ON, OE, OW & -\\
\hline
\end{tabular}
\label{TAB:obs}
\end{table}
% - - - - - - - - - - - - - - - - - - - - - - - - - - - - 

%- - - - - - - - - - - - 
\subsection{Scheduling}
%- - - - - - - - - - - - 
The schedules were prepared with the {\it sked} software \citep{Gipson2010}.
We added the OTT to the necessary catalogues,  e.g. the system equivalent flux density (SEFD) values.
For OTT, we used conservative X-band SEFD values of 3000~Jy, as estimated during the commissioning phase.
The corresponding values for the ON system in the sked-catalogue is 2000~Jy.
We adopted the radio source catalogue that is routinely used in the IVS operational VGOS series (VO).
The schedules were prepared with a minimum elevation cutoff of 5~degree.

ON is one of the S/X telescopes in the IVS with medium-high slew speed, i.e. 3 degree/s in azimuth and 1 degree/s in elevation.
The OTT are fast VGOS-class telescopes with 12 and 6 degree/s in azimuth and elevation, respectively.
To simplify the scheduling, we used a minimum scan length of 30~s and forced the scheduling software to always schedule all three telescopes together.
As a result, the majority of the scheduled scans were 30~s long, and only very few scans were longer. The exceptions were the sessions ON9114, which conservatively used a minimum scan length of 60~s as part of early testing, and ON0318, which was scheduled to explore the impact of more scans using a minimum duration of 10~s. ON0318 did achieve 25\% more scans than the similar ON0317, but these additional observations did not significantly improve the final analysis. 

\begin{figure*}[b!]
\centering
\includegraphics[width=1\textwidth]{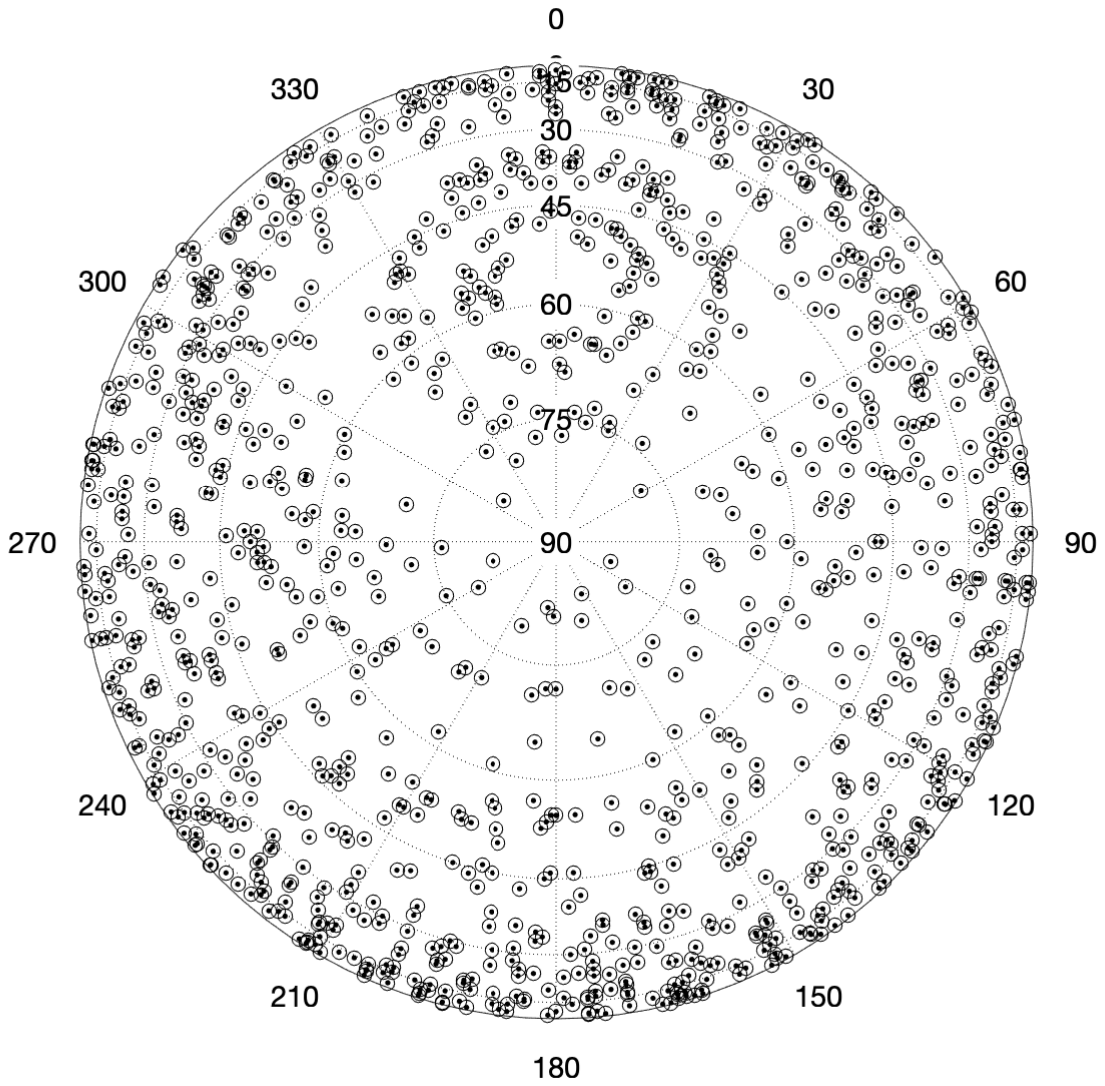}
\caption{Example of a sky plot during a 24~h long ONTIE-experiment at the Onsala Space Observatory. 
The three systems ON-OE-OW always observe together and achieve in this particular example in total 1157 scans, with a total number of 127 different radio sources that were observed.}
\label{FIG:skyplot}      
\end{figure*}

Early tests only covered a few hours of observing time, but once we had gained enough experience, we scheduled the ONTIE-sessions for 24~h or even several consequtive days. The long sessions allow analysis on a range of different timescales.
During a typical 24~h long session, about 1100 scans were scheduled during with more than 125 different radio sources  observed, several of these up to 20 times.
Figure~\ref{FIG:skyplot} shows as an example a typical sky plot obtained during a 24~h long ONTIE-experiment.

%- - - - - - - - - - - - - - - - - - - - - - - - - - - - - - - - -
% FIG-03
%- - - - - - - - - - - - - - - - - - - - - - - - - - - - - - - - -
%- - - - - - - - - - - - - - - - - - - - - - - - - - - - - - - - -

% - - - - - - - - - - - - - - - - - - - - - - - - - - - - 
\subsection{Polarisation and frequency setup}
% - - - - - - - - - - - - - - - - - - - - - - - - - - - - 
The ON system was using right circular polarization (RCP; hereafter referred to as R), while the OTT systems were using linear dual-polarization (hereafter referred to as X and Y). 
The combination of circular and linear polarisation basis, sometimes referred to as mixed-mode, is not a standard mode of observing and limited support exist for this in various software packages. 
Mixed-mode observing was, however, the only available option given the telescope systems involved. Further details regarding mixed-polarisation fringe-fitting are noted in Sect. \ref{sec:fringefitting}.

Various frequency configurations were tried in observations, aiming to avoid local radio frequency interference (RFI), as well as exploring the impact on the final analysis. 
Table~\ref{TAB:freqs} lists the frequency configuration used for all experiments included in this work. Initially, we simply used the standard frequency setup for IVS-R1 sessions with the ON DBBC2 backend \citep{Tuccari_DBBC2_2010}, focusing on 8 channels of 16~MHz bandwidth between 8213.99~MHz and 8949.99~MHz (configuration A in Table~\ref{TAB:freqs}). 
At the time, 16~MHz was the maximum available bandwidth for ON (DBBC2 firmware v105 limitation), while the OTT (with DBBC3 backends) observed with 32~MHz bandwidth per channel. Table~\ref{TAB:freqs} shows the final correlated overlapping bandwidth used in the analysis. 
Starting with experiment ON9135, the DBBC2 firmware was  upgraded to v107 which allowed 32~MHz channels also for ON. 
This allowed us to double the bandwidth and thereby increase the signal-to-noise (configuration~B). 

Configuration~C was an attempt to avoid RFI in the lowest channels a,b which were often excluded from the analysis of data using configuration~B (see Table~\ref{TAB:obs}). 
Unfortunately RFI appeared also in a new channel~e, but we decided to stick with configuration~C for all later experiments as it spanned a larger total frequency range than configuration~B.

For all experiments, ON frequency channels were recorded as upper sideband (USB) while OE and OW data were recorded as lower sideband (LSB). 
This corresponds to the standard (S/X and VGOS) backend configurations for the respective systems.

% - - - - - - - - - - - - - - - - - - - - - - - - - - - - 
% TAB-
% - - - - - - - - - - - - - - - - - - - - - - - - - - - - 
\begin{table}[ht!]
\centering
\caption{List of frequency configurations for all experiments listed in Table \ref{TAB:obs}. The frequencies denote the lower-edge of each correlated BBC-channel in MHz with the bandwidth, 16 or 32 MHz, given in parenthesis.}
\begin{tabular}{ l | l | l | l  }
\hline
Configuration &  A &  B &  C \\ 
Fourfit channel &  &  &  \\ 
\hline
a & 8213.99 (16) & 8212.99 (32)&  8244.99 (32)\\
b & 8253.99 (16) & 8252.99 (32)&  8284.99 (32)\\
c & 8353.99 (16) & 8352.99 (32)&  8384.99 (32)\\
d & 8513.99 (16) & 8512.99 (32)&  8544.99 (32)\\
e & 8733.99 (16) & 8732.99 (32)&  8764.99 (32)\\
f & 8853.99 (16) & 8852.99 (32)&  8884.99 (32)\\
g & 8873.99 (16) & 8892.99 (32)&  8924.99 (32)\\
h & 8933.99 (16) & 8932.99 (32)&  8964.99 (32)\\
\hline
\end{tabular}
\label{TAB:freqs}
\end{table}
% - - - - - - - - - - - - - - - - - - - - - - - - - - - - 

%- - - - - - - - - - - - - - - - - - - - -
\subsection{Phase- and cable calibration}
\label{SEC:pcalintro}
%- - - - - - - - - - - - - - - - - - - - -

\begin{figure*}[b!]
\centering
\includegraphics[width=1\textwidth]{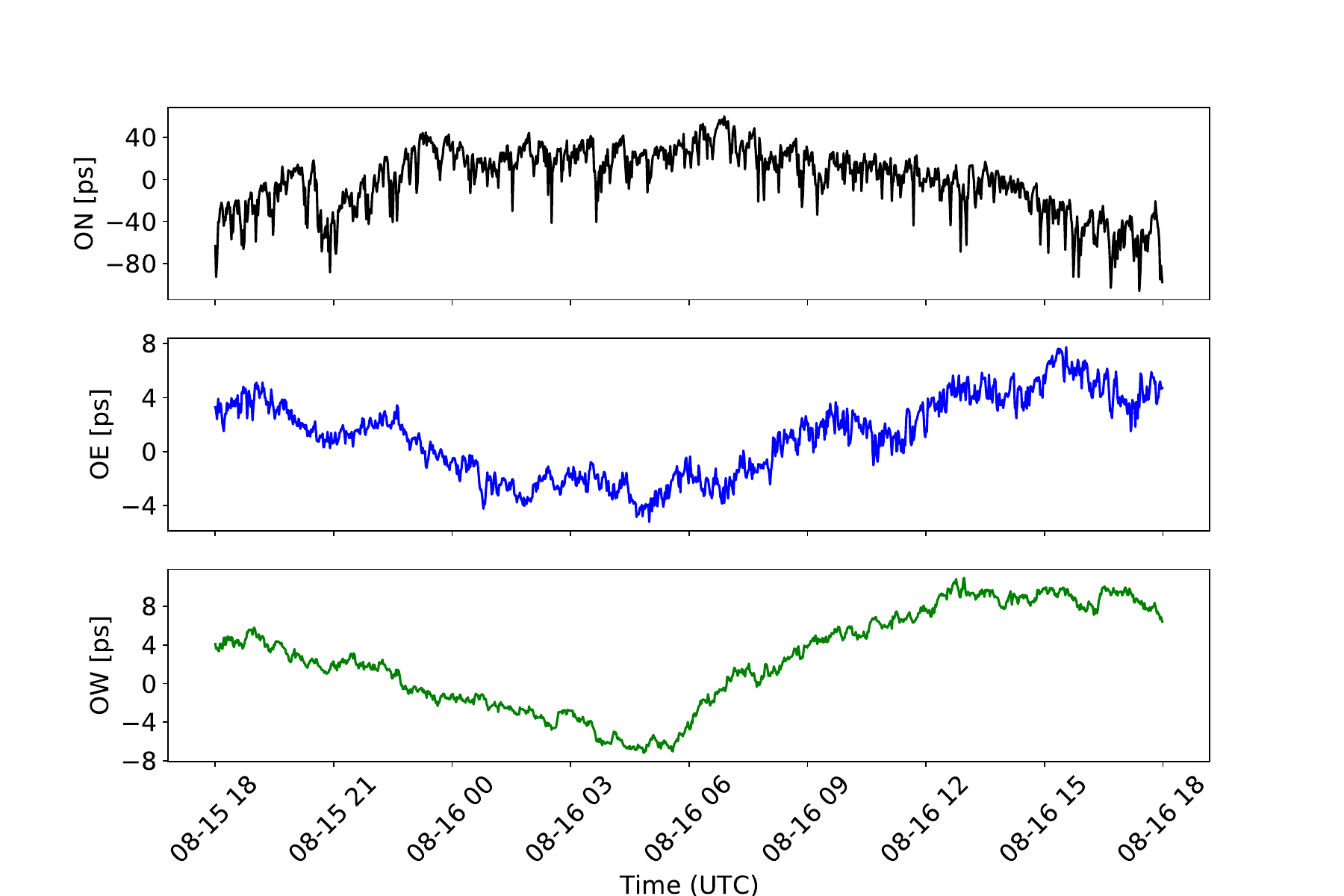}
\caption{Cable delay calibration data for ON (top), OE (middle) and OW (bottom) for session 20AUG15VB. 
Note that the values for ON are one order of magnitude larger than for OE and OW.}
\label{fig:cablecal}
\end{figure*}

For standard geodetic VLBI sessions, ON is working with a 1~MHz phase-calibration (PCAL) system, while the OTT are equipped with more modern 5~MHz PCAL systems. 
It is well known that ON suffers from rather strong direction dependent cable delay variations (Dan MacMillan, private communication, 2002), which will affect the station position if not correctly compensated for in the analysis. This is usually done by using PCAL signals in the post-processing (fringe-fitting) step, and cable calibration data in the analysis step. We note that ON cable cal was always applied with the same  sign (-), even though some observing logfiles were missing this information. 
An example of cable delay calibration data for the three antennas is presented in Fig.~\ref{fig:cablecal}.

%- - - - - - - - - - - - - - - - - - - - - - - - - - - - - - - - -
% FIG-02
%- - - - - - - - - - - - - - - - - - - - - - - - - - - - - - - - -
%- - - - - - - - - - - - - - - - - - - - - - - - - - - - - - - - -

In general, both PCAL and cable cal are therefore necessary to obtain accurate telescope positions. 
However, the early ONTIE-sessions were observed without PCAL.
The reasons were that a) phase calibration and cable delay measuring system (CDMS) of OW was not working in first part of 2019 and needed to be repaired at Haystack observatory, and b) that we were concerned about how to deal with the potential correlation of the PCAL signals themselves.
In the second half of 2019, the OW system was repaired, and we also learned about the possibility of using notch filters in post-processing to reduce the impact of correlating PCAL-signals. The notch filters were the same for all scans and all antennas for an experiment, but changed slightly between experiments to account for RFI differences. The exact details are given in the fourfit control file for each experiment, included in the correlator report available directly via the IVS, and  also included in the respective vgosDb.
Therefore, PCAL was enabled in the ONTIE-sessions starting with ON9323. 
We note that the ON PCAL/cable system was malfunctioning during experiment ON0223, but was repaired for ON0227.

%- - - - - - - - - - - - - - 
\subsection{Data recording}
%- - - - - - - - - - - - - -
The data were recorded using jive5ab \citep{jive5abweb} on the Onsala flexbuff computers. 
ON data were sampled using a DBBC2 backend and recorded as single-thread 8-channel 2-bit sampled VDIF files. 
For the early 16~MHz ONTIE-sessions, this meant 512~Mbps data rate, while for the later 32~MHz experiments it meant 1~Gbps data rate. 

OE and OW data were sampled using two DBBC3 units and, in early ONTIE-sessions, recorded as 8-thread 8-channel  2-bit VDIF data, at 2~Gbps. 
The multi-thread data were converted to single-thread 64-channel data using the tool \emph{vmux} (bundled with DiFX) for efficient correlation. 
This worked, but delayed correlation due to the extra processing time and space needed for the conversion process. 
In later experiments, jive5ab version 3.0 was used which allowed recording the 8 streams from each OE/OW DBBC3 into 8 separate single-thread VDIF files.
These could be correlated directly without \emph{vmux}, allowing quicker processing of the data.

%- - - - - - - - - - - - - - - - - - - - - - - - - - - - - - - - -
\section{Correlation and post-correlation processing}
\label{SEC:correlation}
%- - - - - - - - - - - - - - - - - - - - - - - - - - - - - - - - -
In this section we described in detail the correlation of the VDIF data, as well as the post-processing steps applied after correlation to obtain vgosDbs for geodetic analysis.

%- - - - - - - - - - - - - - - - - - 
\subsection{Correlation using DiFX}
%- - - - - - - - - - - - - - - - - -
The OTT systems are equipped with DBBC3 \citep{Tuccari_DBBC3_2018} backends which, in the current firmware used for VGOS observations (v123 or 124), only allows a channel width of 32~MHz. 
%(Other bandwidths will be allowed in the upcoming firmware version 125). 
This meant a bandwidth mismatch between ON (16~MHz) and OTT (32~MHz) in initial experiments (of which ON9114 and ON9120 are included in this work). 
To overcome this, we used the \emph{zoomband} capability of the DiFX software correlator \citep{Deller2007, Deller2011} to correlate matching 16~MHz channels on all baselines. 
In theory, we could correlate 32~MHz channels on the OE-OW baseline and 16~MHz on the ON-OE/OW baselines. While this could improve the signal-to-noise  on the OE-OW baseline, we opted for 16~MHz everywhere in the early experiments for simplicity. 

Starting with experiment ON9135, all telescope recorded 32~MHz channels and therefore we correlated these with 32~MHz channel width on all baselines. 
Surprisingly, we found it necessary to use \emph{zoomband} option in DiFX also for the 32~MHz channels (although all channels matched perfectly). If not, no correlation products were obtained for the OE-OW baseline. After discussions with DiFX developers, we believe this happened because ON record USB and OE/OW record LSB. This causes DiFX to generate two frequency tables, but only one table is correctly interpreted by difx2mark4, hence losing the LSB-LSB baseline (OE-OW). We will investigate this further together with DiFX developers, but as we did not find any negative effects using zoomband we do not see this limiting the analysis presented in this paper.

The data were correlated on one of the Onsala flexbuff computers using DiFX release version 2.6.1. 
For data-logistic reasons different machines were used to correlate different experiments, but all running identical software. 
On a typical machine (12 CPUs á 4.6~GHz, 128~GB RAM, 36x12~TB storage) the correlation wall-time was about 1:1 compared to the observing time (when excluding the vmux-step necessary for early multi-thread experiments). 
The spectral resolution used for correlation was 0.25~MHz (corresponding to 128 lags per 32~MHz channel bandwidth).

%- - - - - - - - - - - - - - - - - - - - - - -
\subsection{Fringe-fitting with HOPS fourfit}
\begin{figure*}[ht!]
\centering
\includegraphics[width=\textwidth]{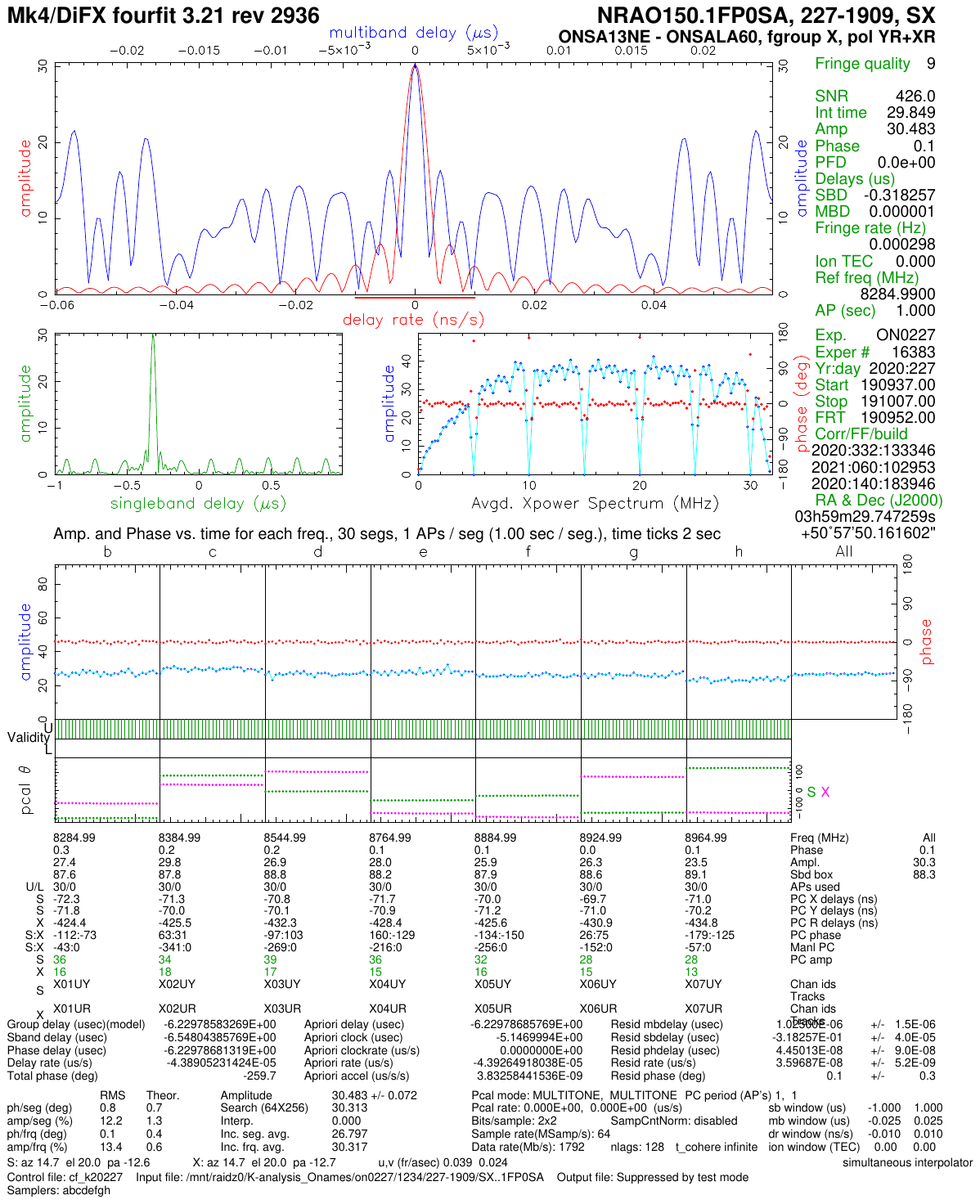}
\caption{Example of fringes obtained during session ON0227. 
This is a fourfit plot showing one scan on the radio source NRAO150 between OE and ON. 
To maximise the signal-to-noise for the geodetic analysis, the mixed-basis (RX,RY) correlation products are added together in the fringe-fitting as YR+XR. 
The regular drops in amplitude every 5~MHz correspond to channels with cross-correlating phasecal signals which have been excluded from fringe-fitting using \emph{notches} in fourfit.}
\label{fig:fringes}
\end{figure*}

\begin{figure*}[ht!]
\centering
\includegraphics[width=1\textwidth]{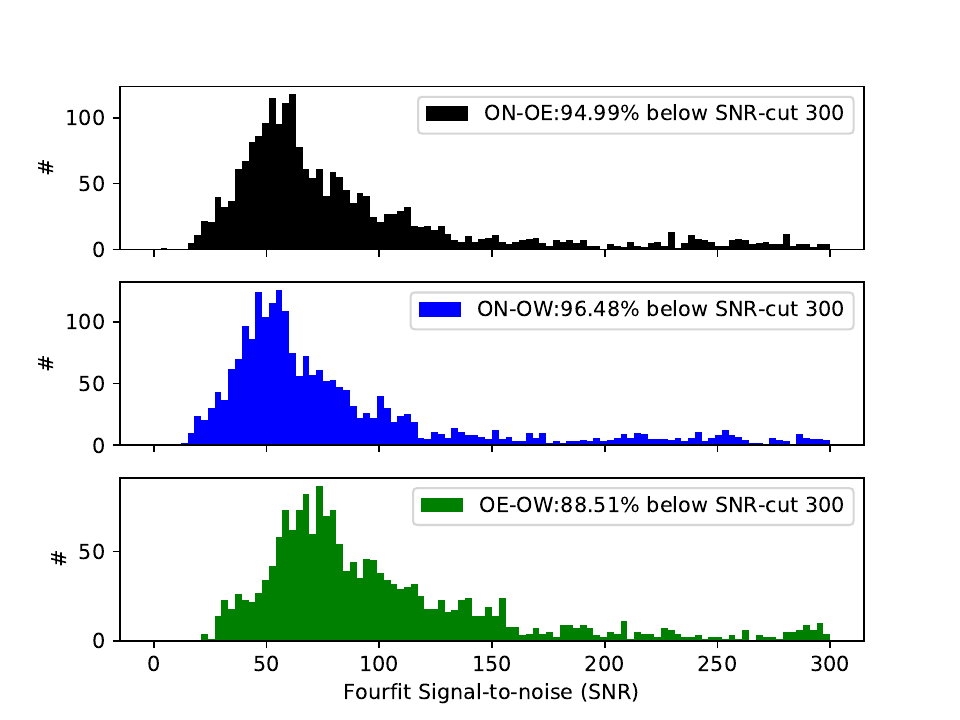}
\caption{Histogram of fourfit SNR for all observations without G-codes in experiment ON0227. 
Values greater than 300 have been excluded for clarity, and the legend shows the percentage of (non-G-code) observations shown. 
It is clear that the OE-OW baseline (bottom) is more sensitive than the ON-OE/OW baselines.}
\label{fig:SNR}
\end{figure*}

%- - - - - - - - - - - - - - - - - - - - - - -
\label{sec:fringefitting}
The post-correlation processing was done with the latest stable version 3.21 of the Haystack Observatory Postprocessing System (HOPS) \citep{HOPSweb}. 
To achieve proper PCAL extraction for multi-stream VDIF, version 1.6 of {\it difx2mark4} was used to convert from SWIN (difx output) to MK4 (fourfit input) format. 
Fringe fitting was done using HOPS {\it fourfit} where, as standard geodetic analysis-software is only able to process one polarisation product per baseline, pseudo-stokes I was formed on the OE-OW baseline, and RX+RY on the ON-OE/OW baselines.  
An example result, showing one scan of the source NRAO150 on baseline ON-OE observed in experiment ON0227, is shown in Figure~\ref{fig:fringes}. 
This one of the brightest sources; a histogram of the SNR obtained on the three baselines in experiment ON0227 is presented in Figure~\ref{fig:SNR}.

%- - - - - - - - - - - - - - - - - - - - - - - - - - - - - - - - -
% FIG-04
%- - - - - - - - - - - - - - - - - - - - - - - - - - - - - - - - -
%- - - - - - - - - - - - - - - - - - - - - - - - - - - - - - - - -

%- - - - - - - - - - - - - - - - - - - - - - - - - - - - - - - - -
% FIG-05
%- - - - - - - - - - - - - - - - - - - - - - - - - - - - - - - - -
%- - - - - - - - - - - - - - - - - - - - - - - - - - - - - - - - -

For antennas lacking PCAL and/or cable calibration during a significant part of the experiment, manual PCAL was used in fringe-fitting. 
In this case, channel based phase corrections were determined using one scan on a bright source (often NRAO150), with \verb!pc_mode manual! being used in fourfit. 
ON R-pol was used as reference, i.e. corrections were found for OE/OW X/Y-pol.
Where PCAL and cable calibration was present, \verb!pc_mode multitone! PCAL was used in fringe-fitting. 
Additional channel-based phase-corrections, on top of multitone PCAL, were determined for the OE/OW X and Y polarisations (ON R-pol) as reported by fourfit (using option \verb!-m 1!) \verb!pc_phases! on a bright source (in most experiments NRAO150). 

Because all three telescopes PCAL systems are tied to the same hydrogen maser, the PCAL signals correlate strongly. To reduce the impact of correlating PCAL signals when fringe-fitting, \verb!notches! were used in fourfit every 5~MHz to remove the PCAL signals of the OTT. The notch filters removed $\pm0.5$~MHz of bandwidth around the PCAL tones, significantly reducing the impact of correlating PCAL signals on fringe-fitting. This causes regular drops in the cross-power spectra, also visible in Figure~\ref{fig:fringes}. Minor residuals, resulting from the periodic structure of the removed PCAL signals, can be seen in the singleband delay panel in Figure~\ref{fig:fringes}. For clarity we note that the PCAL data, extracted by DiFX, were  available to fourfit for use in fringe-fitting. The notch filters just removed the problematic channels from the visibility data, in the same way that one can mask out correlating narrow-band RFI. Channels completely removed due to broadband RFI are listed in Table~\ref{TAB:freqs}.

%- - - - - - - - - - - - - - - - - - - -
\subsection{Creation of VGOS databases}
\label{SEC:vgosdb}
%- - - - - - - - - - - - - - - - - - - -
After fringe-fitting, the obtained VLBI delay observations were exported to the so-called vgosDb format \citep{Bolotin_et_al_2015} that can be read by standard geodetic VLBI analysis software packages. 
Three database \emph{wrappers} were created using utilities bundled with the $\nu$Solve software package version 0.7.1 \citep{Bolotin_et_al_2012}. 
Wrapper~1 was created from fringe-files using \emph{vgosDbMake 0.5.1}. 
In this step we decided to split observations spanning several days into separate databases with approximately 24~h in each database: see Table \ref{TAB:obs}. 
Wrapper~2 was created using \emph{vgosDbCalc 0.4.1}. 
Finally, Wrapper 3 was created using \emph{vgosDbProcLogs 0.5.1} which appended relevant supplemental data from the observing logfiles (temperature, pressure, humidity etc.). 
In this final step, we used pressure-corrected logfiles as described in Sect.~\ref{SEC:pcor}.

\subsection{Editing and ambiguity resolution}
The version 3 wrappers were now read by $\nu$Solve, and processed in a standard way to remove obvious outliers and resolve group-delay ambiguities. The results were written to a new wrapper version 4. Finally, we used $\nu$Solve (version 0.7.3; which can save phase-delay ambiguity information) to resolve phase-delay ambiguities and store the result in wrapper version 5.

%- - - - - - - - - - - - - - - - - - - - - - - - - - - - - - - - -
\section{Geodetic data analysis}
\label{SEC:analysis}
%- - - - - - - - - - - - - - - - - - - - - - - - - - - - - - - - -
We used the software package {\em ASCOT} {(git commit hash daa99c2)\footnote{Available via \url{https://github.com/varenius/ascot}. }} \citep{Artz_et_al_2016} for our geodetic analysis. 
{\em ASCOT} is a fully developed geodetic VLBI analysis software package and has been used recently for the contribution of IVS Analysis Center at the Onsala Space Observatory for its contribution to the ITRF2020 calculations\footnote{\url{ https://ivscc.gsfc.nasa.gov/IVS_AC/IVS-AC_ITRF2020.htm}}.
The software is able to model both thermal \citep{Nothnagel_2009} and gravitational \citep[see e.g.][]{Nothnagel_et_al_2019,Loesler_et_al_2019} deformation of radio telescopes, which has a significant impact on the local-tie vectors (see Sect.~\ref{SEC:gravdef}). 
{\em ASCOT} is also capable of analysing both group-delay and phase-delay observations, and combine the results from multiple databases into one \emph{global solution}.

%- - - - - - - - - - - - - - - - - - - - 
\subsection{A~priori data and modelling}
\label{SEC:apriori}
%- - - - - - - - - - - - - - - - - - - - 
{\em ASCOT} (and \ns{}) make use of a~priori data, such as initial station positions, site velocities and Earth Orientation Parameters (EOP). 
Some of these parameters may have a significant impact on the final results, and therefore we strive to use the best available a~priori information.

The ON position was fixed to the value presented in VTRF2020b \citep{VTRF2020b}. We note that this is the first VTRF release which includes gravitational deformation modeling for ON \citep{Thaller_2021}.
Initial positions for OE and OW were based on Real-time Kinematic (RTK) GPS measurements carried out in 2016, plus information taken from the construction drawings. 

All three stations were assumed to have velocities identical to the values for ONSALA60 in VTRF2020b  \citep{VTRF2020b}. 
We used \emph{axis offsets} of 0~mm for OE, OW and $-6$~mm for ON \citep{Haas_Eschelbach_2005}. 
EOP information was taken from the most recent IERSC04, and the VMF3 tropospheric mapping function \citep{VMF3} was used in all {\em ASCOT} processing.

Source positions of the quasars observed were assumed as their ICRF3 S/X values \citep{charlot2020}. \footnote{Available via http://hpiers.obspm.fr/icrs-pc/newwww/icrf/icrf3sx.txt.}

Corrections for cable delay variations (CDMS for OTT) were applied in the analysis if, and only if, the antennas had working PCAL which could be used in fringe-fitting. 
For antennas labelled with \emph{Manual PCAL} in Table~\ref{TAB:obs}, no cable calibration was applied when analysing the particular vgosDb.

%- - - - - - - - - - - - - - - - - - - - -
\subsubsection{No ionosphere corrections}
%- - - - - - - - - - - - - - - - - - - - -
\label{SEC:ion}
In standard dual- and multi-frequency VLBI observations, ionospheric corrections are usually applied in the analysis. 
Such corrections were not possible to apply in our case since we only observed X-band. 
However, the three stations ON, OE and OW are located within about 550~m distance, i.e. the differential ionospheric effects are negligible. 
Ionospheric corrections could thus be omitted without any significant impact on the analysis.

%- - - - - - - - - - - - - - - - - - - - - - - - - - - - -
\subsubsection{Pressure correction of observing log files}
%- - - - - - - - - - - - - - - - - - - - - - - - - - - - -
\label{SEC:pcor}
All three telescope systems at OSO use the same meteorological station, and the corresponding data are recorded identically in the three individual logfiles. 
However, the ellipsoidal height of this meteosensor is not identical to the ellipsoidal height of neither the reference point (intersection of telescope's azimuth and elevation axes) of ON, nor of OTT.

We therefore corrected the pressure readings as recorded in log files correspondingly, using a height-dependent pressure correction \citep{Berg_1948}. 
The corrected pressure $P_n$ is calculated according as
%- - - - - - - - 
\begin{equation}
      P_n = P_r \times (1 - 0.0000226\times(h - h_r))^{5.225}
      \label{eqn:p}
\end{equation}
%- - - - - - - - 
where $P_r$ is the original field-system wx log value, $h$ is the height of ON (59.3~m), OE (53.2~m) or OW (53.2~m), and $h_r$ is the height of the OSO pressure sensor (46.6~m). 
The approximate changes with respect to the original observing log-values are ON = $-1.5$~hPa, OE = $-0.8$~hPa, OW = $-0.8$~hPa.
Since the ellipsoidal heights can be assumed to be accurate within 10~cm, the pressure corrections are expected to be accurate within 0.1~hPa. 
Using the corrected pressure values in the vgosDbs ensure that the zenith hydrostatic delays can be modelled as good as possible. 
We note that the log files available via the IVS are the corrected logfiles, with the pressure correction already applied.

\subsubsection{Analysis settings and parametrisation}
All observed scans were correlated (no data loss), and all correlated scans were post-processed. A few observations were removed in the geodetic analysis, either due to low quality codes from fringe-fitting, or due to being significant outliers. The numbers vary between the vgosDbs, but in general $>95\%$ of all scheduled observations were used in the final analysis.

In addition to OE and OW station positions, we modelled ZWD (every 30 minutes), clocks (every 60 minutes), thermal expansion of the antennas, and gravitational deformation of the antennas. The significance of these effects are investigated in Sect. \ref{sec:modeling}.

We also added 3~ps and 1.5~ps of extra elevation-dependent noise to the group- and phase-delay observations respectively. 
Addition of extra noise is standard procedure in geodetic VLBI analysis to account for over-optimistic weights in the delay observations, and hence bring the reduced $\chi^2$ closer to 1 \citep{Gipson_2007}.

\subsection{Group-delay analysis}
\begin{figure*}[htb]
\centering
\includegraphics[width=1\textwidth]{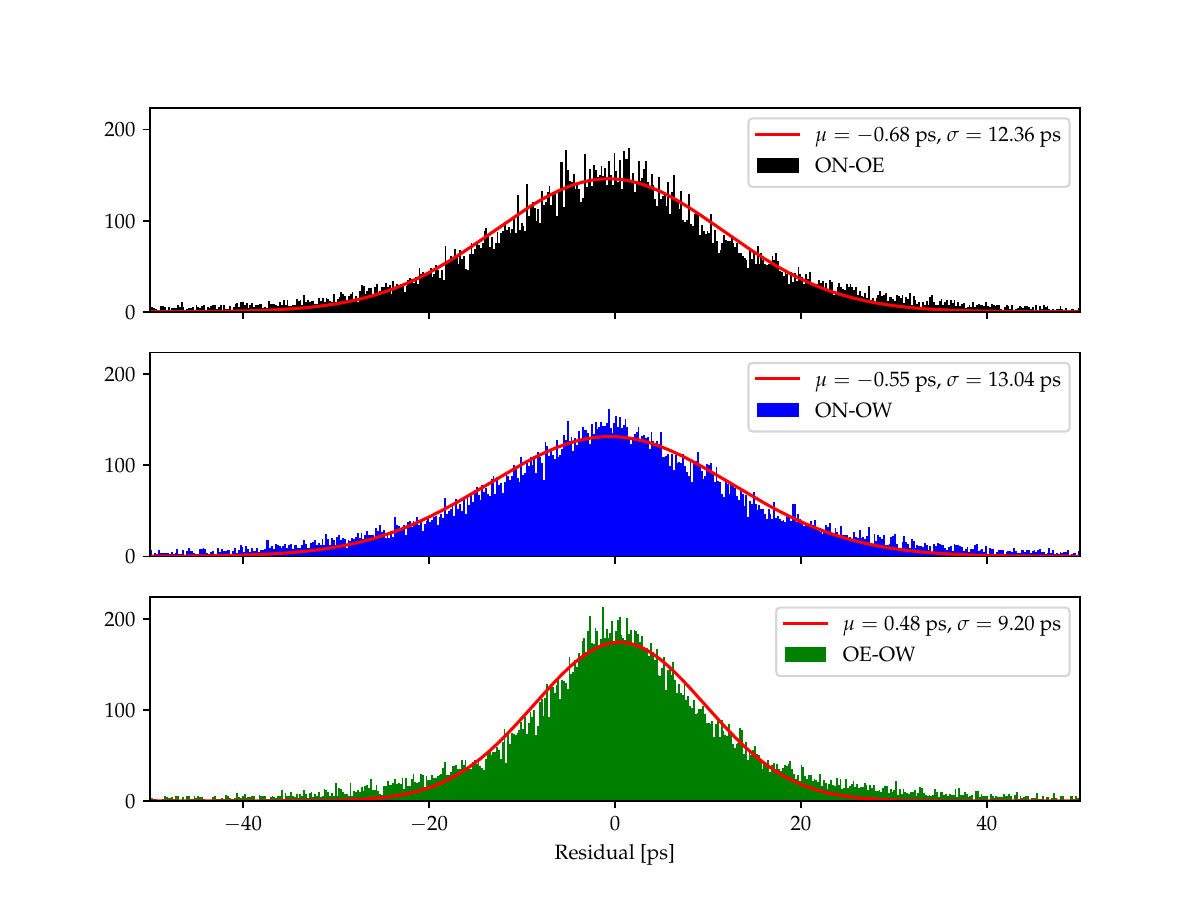}
\caption{Histograms of postfit group-delay residuals of all 25 ONTIE-sessions.
The red lines are the corresponding fitted normal distributions, and their mean values $\mu$ and standard deviations $\sigma$ are given. }
\label{FIG:GR_HIST}      
\end{figure*}

Using {\emph ASCOT} we obtain, for each vgosDb in Table~\ref{TAB:obs}, group-delay position estimates for OE and OW. The OE and OW group-delay positions for all vgosDbs are presented in Tables \ref{TAB:gr-oepos} and \ref{TAB:gr-owpos} respectively. To illustrate the results, we present in Figure~\ref{FIG:GR_HIST} three histograms of the post-fit residuals on the three baselines from the group-delay analysis of all ONTIE-sessions with {\em ASCOT}. Normal distributions are fitted to these histograms and show that the mean residuals are below 1~ps on all three baselines, with standard deviations below 15~ps.
As expected the short baseline between the modern VGOS antennas OE and OW gives the best performance.
%- - - - - - - - - - - - - - - - - - - - - - - - - - - - - - - - -

\begin{figure*}[htbp]
\centering
\includegraphics[width=1\textwidth]{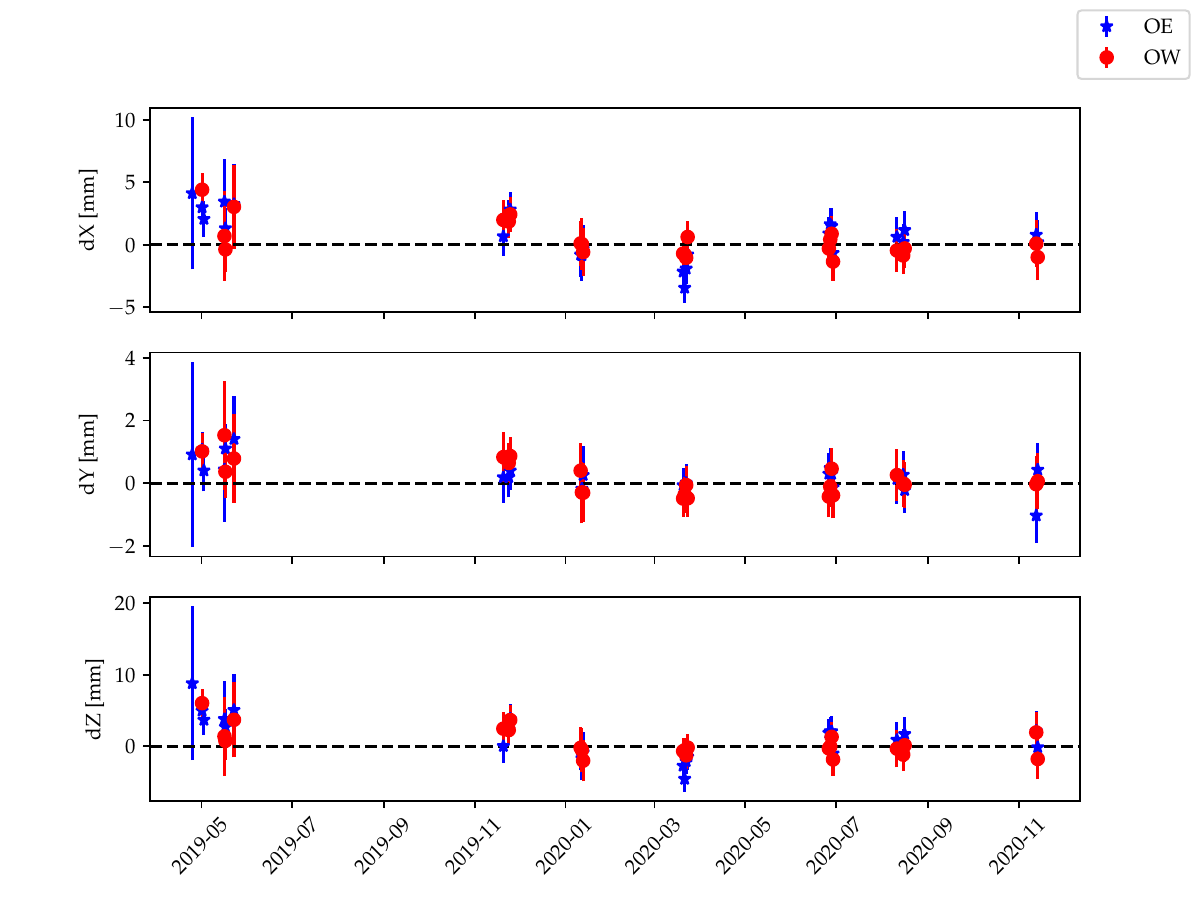}
\caption{Group-delay positions for for OE (blue stars) and OW (red circles), with 3$\sigma$ uncertainties. The final global group-delay positions have been subtracted. Except for a few early points (without phase- and cable calibration available), the individual results are in good agreement with the global position.}
\label{fig:GR_vs_time}
\end{figure*}

Another way to illustrate the results is to look at the OE and OW positions obtained as a function of time, i.e. from each vgosDb. A comparison of the group-delay positions for the 25 ONTIE-session are shown in Figure \ref{fig:GR_vs_time}.

%\subsubsection{Global group-delay positions}
\subsubsection{Station positions derived from global analysis of group-delays}
\label{SEC:globalGR}
{\emph ASCOT} allows so called \emph{global solutions}, where multiple sessions are combined to estimate one set of coordinates for the whole time span. This is done by stacking the normal equations of the individual sessions into one big normal equation system. By inverting this combined system of normal equations, we obtained the station coordinates of OE and OW, as well as the associated variance-covariance matrix. From the latter we then calculated the formal uncertainties of the positions. In  this combined solution, we only included sessions which had phase- and cable calibration applied for all antennas (see Table \ref{TAB:obs}). The resulting global (group-delay) OE and OW positions are given in Table \ref{TAB:grpos-avg}.

\subsection{Phase-delay analysis}
\begin{figure*}[htb]
\centering
\includegraphics[width=1\textwidth]{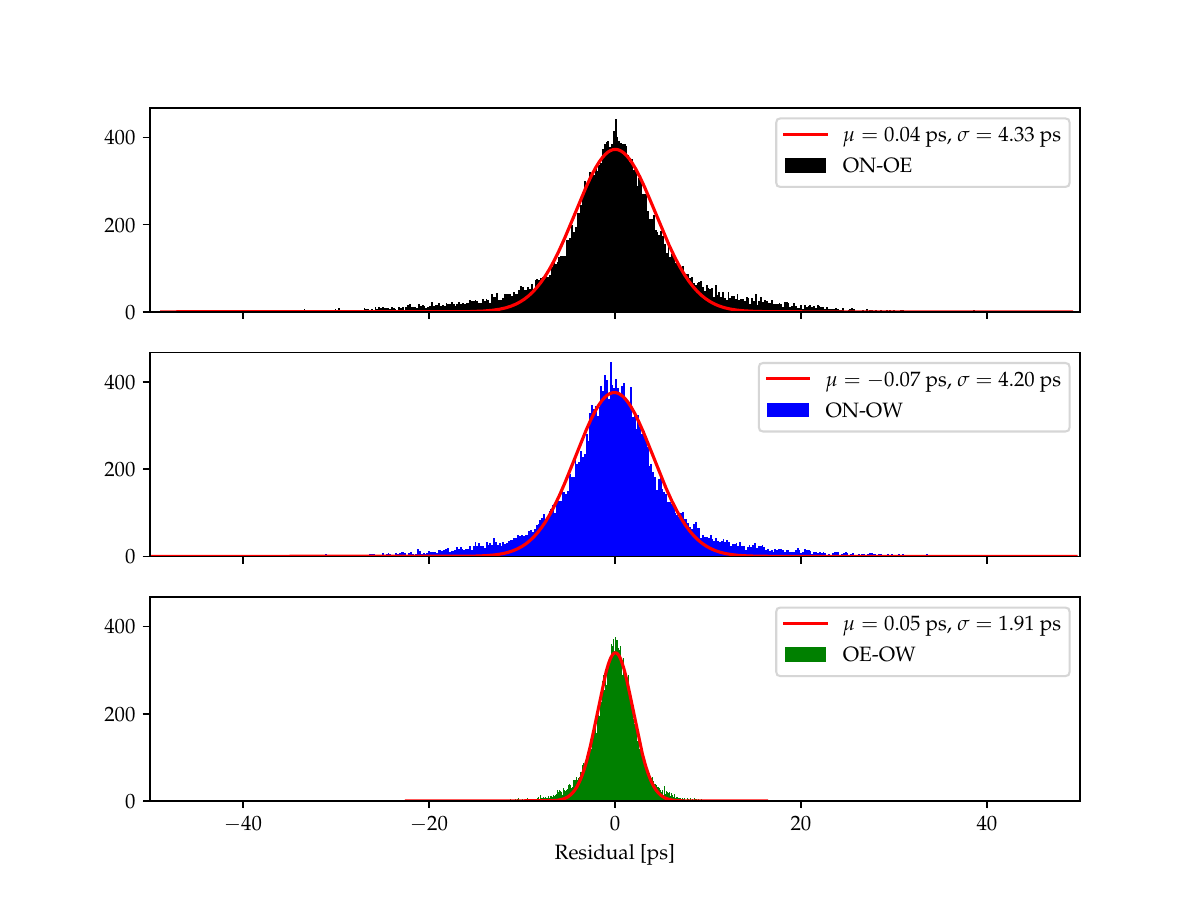}
\caption{Histograms of postfit phase-delay residuals of all 25 ONTIE-sessions.
The red lines are the corresponding fitted normal distributions, and their mean values $\mu$ and standard deviations $\sigma$ are given. For easy comparison, the same horizontal axis range is used as in Figure \ref{FIG:GR_HIST}.}
\label{FIG:PH_HIST}
\end{figure*}

With {\emph ASCOT} we also obtain, for each vgosDb in Table~\ref{TAB:obs}, phase-delay position estimates for OE and OW. The OE and OW phase-delay positions for all vgosDbs are presented in Tables \ref{TAB:ph-oepos} and \ref{TAB:ph-owpos} respectively.
Histograms of all phase-delay residuals are presented in Figure~\ref{FIG:PH_HIST}, with mean residuals below 0.1~ps and standard deviations below 5~ps. This is, as expected, a significant improvement in the uncertainties with respect to the group-delay results. Similarly to our group-delay results, the short baseline between the modern VGOS antennas OE and OW gives the best performance, also for phase-delays.

\begin{figure*}[htbp]
\centering
\includegraphics[width=1\textwidth]{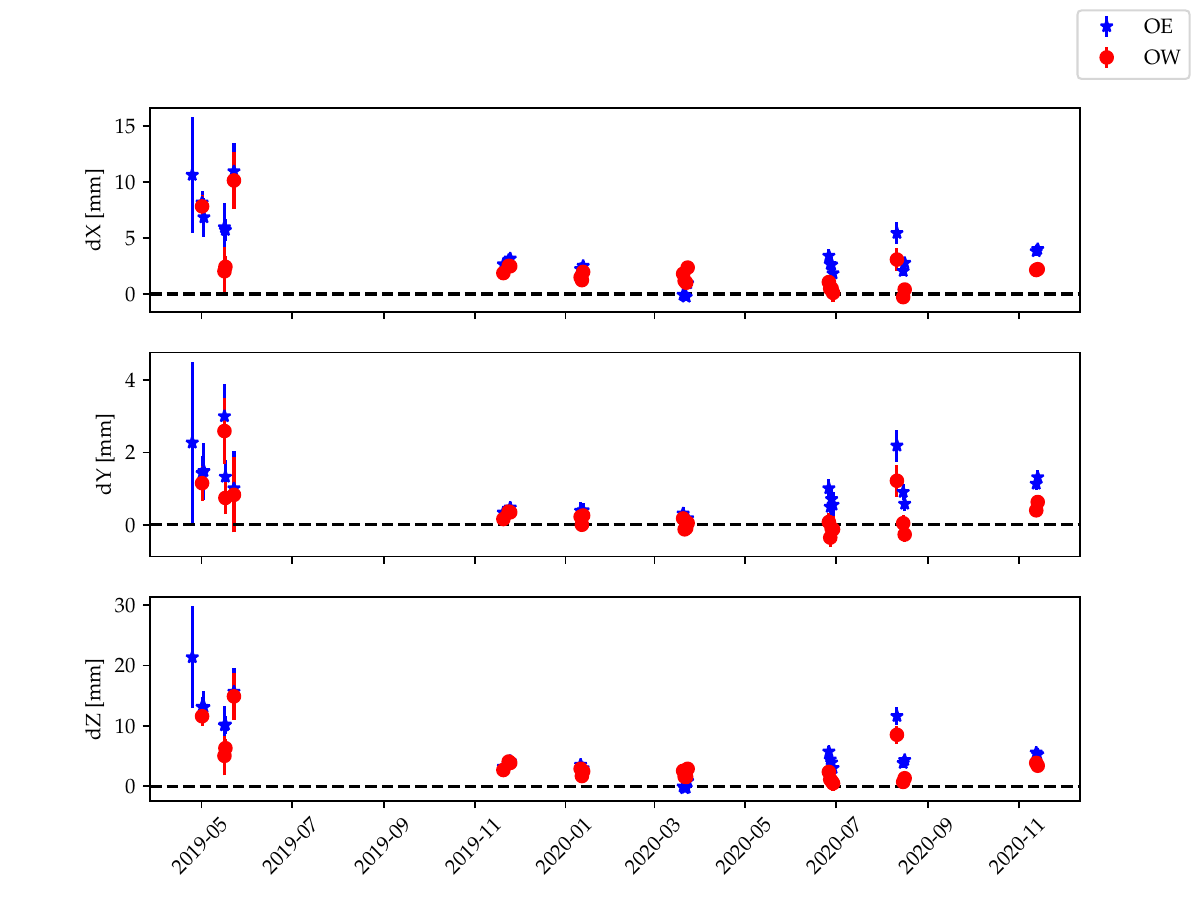}
\caption{Phase-delay positions for for OE (blue stars) and OW (red circles), with 3$\sigma$ uncertainties. The final global group-delay positions have been subtracted. Two significant offsets can be seen: a) about 1~cm in X and Z between experiments with/without cable- and phase-cal, and b) a systematic offset in X and Z ($\approx$3~mm in the Up-direction; see also Table \ref{TAB:grphdiff}) between the phase-delay and group-delay positions.} 
\label{fig:PH_vs_time}
\end{figure*}

\subsubsection{The effect of phase- and cable calibration}
As noted in Sect.~\ref{SEC:pcalintro}, ON suffers from direction dependent cable delay variations (Dan MacMillan, private communication 2002), which will affect the observed delays if not correctly compensated for in the analysis. Indeed, we find a significant systematic offset of about 1~cm in X and Z (see Figure~\ref{fig:PH_vs_time}) between experiments with/without PCAL and cable corrections for ON (listed in Table \ref{TAB:obs}).

%\subsubsection{Global phase-delay positions}
\subsubsection{Station positions from global analysis of phase-delays}
\label{SEC:globalPH}
As with the group-delay data, we can use {\emph ASCOT} to obtain a {\it global solution} of phase-delay data, i.e. a combined phase-delay solution for multiple sessions. 
Again, we use the sessions with phase- and cable calibration present, and we obtain the OE and OW positions presented in Table \ref{TAB:phpos-avg}.

\begin{table}[htb]
\centering
\caption{VTRF2020b (epoch 2010.0) phase-delay positions (in m) and their formal standard deviations (in mm) for OE and OW, obtained as described in Sect. \ref{SEC:globalPH}.}
\begin{tabular}{ l | c | c | c | c | c | c}
antenna & X  & Y   & Z   & $\sigma_X$ & $\sigma_Y$ & $\sigma_Z$ \\
\hline
OE & 3370889.29933 & 711571.19930 & 5349692.05005 & 0.04 & 0.02 & 0.06 \\
OW & 3370946.77978 & 711534.50658 & 5349660.92639 & 0.04 & 0.02 & 0.06 \\
\hline
\end{tabular}
\label{TAB:phpos-avg}
\end{table}

\subsection{Differences in group-delay and phase-delay positions}
Contrary to expectations, we find a significant systematic offset between the group- and phase-delay positions. This can be seen in  Figure~\ref{fig:PH_vs_time}, where the zero-level is the global group-delay position. However, the offset becomes clearer after conversion to an ENU system, see Table \ref{TAB:grphdiff}. 
\begin{table}[hbt]
\centering
\caption{ENU difference for the global OE and OW positions obtained with phase-delays compared to group-delays. }
\begin{tabular}{l | c c c }
antenna  & $\Delta E$ (mm) & $\Delta N$ (mm) & $\Delta U$ (mm)\\
\hline
OE & $+0.07\pm0.04$ & $-0.19\pm0.01$ & $+3.77\pm0.28$\\
OW & $+0.11\pm0.04$ & $+0.07\pm0.01$ & $+2.52\pm0.28$\\
\hline
\end{tabular}
\label{TAB:grphdiff}
\end{table}
Since the offset is primarily in the Up-direction, it is consistent with an elevation-dependent delay error. ON is known to suffer from significant ($\sim$5~mm) gravitational deformation effects, which manifest primarily in the Up-coordinate. While this is included in our modeling, this gravitational deformation model was primarily developed and tested using group-delay observations. There could be additional effects which affect phase-delay estimation. Other explanations for this systematic offset are also possible, but a detailed investigation of this offset is beyond the scope of this paper. 
We note, however, that group-delay analysis is the most common method routinely employed by IVS analysis centers to derive station positions. Therefore, to avoid confusion in the community, we also chose to present group-delay estimates as the OE and OW station positions in this paper. The systematic offset will be monitored and further investigated in future observations.

\subsection{Modeling clocks, ZWD, and antenna deformation}
\label{sec:modeling}
In this section we investigate how various effects, which are possible to model in {\em ASCOT}, impact the estimated OE and OW positions.

\subsubsection{Impact of clock parameter interval length}
%- - - - - - - - - - - - - - - - - - - - - - - - - - - - - - - - - - - -
In order to investigate the impact of the clock parameter interval length, we compared the estimated OTT positions using both 1~hour and 20~minutes intervals. 
Using 20~minutes increase the number of parameters to be estimated by a factor of three (compared to 1~h), but did not cause a problem for the analysis due to the large number of observations during the usually 24~hour long ONTIE-sessions, with well above 1000 observations per baseline.

Table~\ref{TAB:investCLK} provides the observed changes in the weighted mean of the estimated station positions for the OTT from all ONTIE-sessions.
We find that station positions are not impacted significantly by the choice of clock parameter interval length.
This finding is confirmed from a similar analysis of individual ONTIE-sessions with {\ns}. Since ON, OE and OW all share the same maser clock, any variations between the three systems are likely due to telescope-specific instrumental instabilities. 

% - - - - - - - - - - - - - - - - - - - - - - - - - - - - 
\begin{table}[ht]
\centering
\caption{Effect on the weighted mean station positions of OE and OW when changing from clock interval length of 1~hour to  20~minutes, expressed in a topocentric east-north-up (ENU) coordinate system.}
\begin{tabular}{l | c c c }
antenna   & $\Delta E$ (mm)     & $\Delta N$ (mm)     & $\Delta U$ (mm) \\
\hline
OE & $-0.023 \pm 0.022$ & $-0.008 \pm 0.017$ & $+0.061 \pm 0.043$\\ 
OW & $+0.023 \pm 0.021$ & $-0.032 \pm 0.019$ & $+0.031 \pm 0.061$\\
\hline
\end{tabular}
\label{TAB:investCLK}
\end{table}
% - - - - - - - - - - - - - - - - - - - - - - - - - - - - 

%- - - - - - - - - - - - - - - - - - - - - - - - - - - - - - - - - - - - 
\subsubsection{Impact of estimating Zenith Wet Delay (ZWD)}
\label{SEC:trop}
%- - - - - - - - - - - - - - - - - - - - - - - - - - - - - - - - - - - -
All three stations are located within 550~m and thus to a large extent share the same common local troposphere.
Furthermore, the ONTIE-sessions were scheduled in a way that all three stations observed each scan together, i.e. the antennas had almost identical azimuth and elevation directions.
However, in principle, small variations in the local troposphere and atmospheric turbulence effects could affect the delays on the three baselines in a differential way.

Since the elevation angles for the three antennas were almost identical, it is not possible to estimate antenna-specific tropospheric parameters for all three antennas.
However, it is possible to estimate \emph{differential} ZWD parameters for OE and OW as piece-wise linear offsets every 20~minutes, while not estimating any tropospheric parameters for ON.
Table~\ref{TAB:investZWD} provides the observed changes in the weighted mean of the estimated station positions for the OTT from all ONTIE-sessions when analysing with and witout estimating ZWD for OE and OW. The changes are expressed in a topocentric east-north-up (ENU) coordinate system.
%While the horizontal components are not affected by more than 50~$\mu$m, we note a significant reduction in the up-components of the antennas, on the level of 1.1~mm to 1.6~mm.
While the horizontal components are not affected by more than 30~$\mu$m, we note a significant reduction in the up-components of the antennas, on the level of 0.3~mm to 0.8~mm.

This shows that estimating differential ZWD is of importance, even for this small network.
The estimated differential ZWD themselves are typically within $\pm 1$~mm, and the values in general follow each other for both stations, with some few exceptions.
The overall scatter is on the order of 0.5-1~mm.
A more detailed investigation on the differential ZWD and their potential importance to sense atmospheric turbulence, is the topic of future investigations.
% - - - - - - - - - - - - - - - - - - - - - - - - - - - - 
\begin{table}[htb]
\centering
\caption{Effect on the estimated station positions of OE and OW, expressed in a topocentric (ENU) coordinate system, when estimating ZWD for OE and OW as piece-wise linear offsets with 30~minute interval length, compared to not estimating ZWD.}
\begin{tabular}{l | c c c }
antenna   & $\Delta E$ (mm)     & $\Delta N$ (mm)     & $\Delta U$ (mm) \\
\hline
%OE & $+0.019 \pm 0.041$ & $-0.030 \pm 0.024$ & $-1.641 \pm 0.266$ \\ 
%OW & $+0.050 \pm 0.035$ & $-0.039 \pm 0.022$ & $-1.150 \pm 0.224$ \\ 
OE & $-0.030 \pm 0.040$ & $-0.006 \pm 0.027$ & $-0.784 \pm 0.261$ \\ 
OW & $+0.006 \pm 0.037$ & $-0.023 \pm 0.026$ & $-0.321 \pm 0.226$ \\ 
\hline
\end{tabular}
\label{TAB:investZWD}
\end{table}
% - - - - - - - - - - - - - - - - - - - - - - - - - - - - 

%- - - - - - - - - - - - - - - - - - - - - - - - - - - - - - - - - - - - 
\subsubsection{Impact of thermal deformation}
\label{SEC:thermdef}
\begin{figure*}[bht]
\centering
\includegraphics[width=1\textwidth]{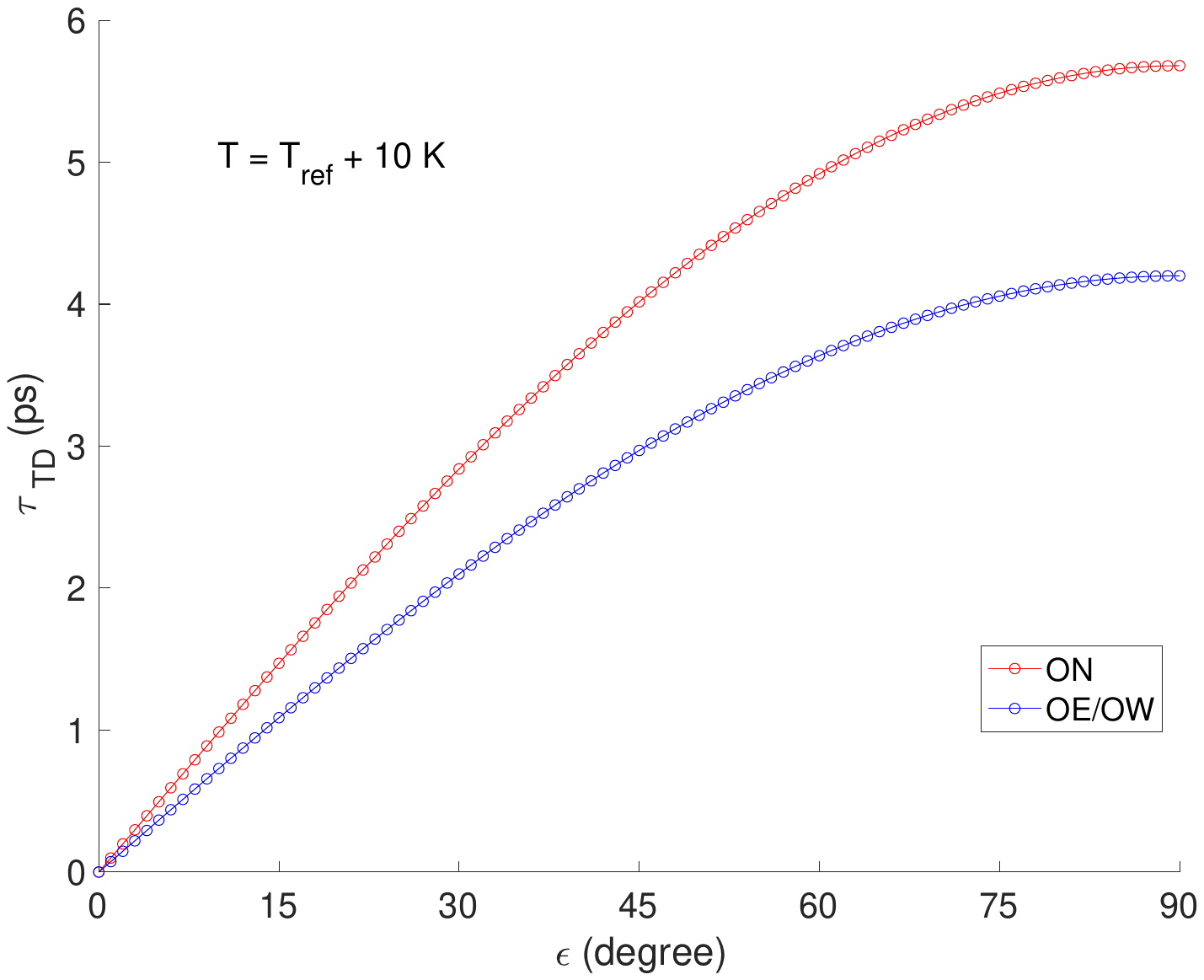}
\caption{Thermal deformation effect on the VLBI delay  \citep{Nothnagel_2009} for the antennas ON (red) and OE/OW (blue).}
\label{FIG:TD}      
\end{figure*}

%- - - - - - - - - - - - - - - - - - - - - - - - - - - - - - - - - - - - 
Figure~\ref{FIG:TD} depicts the expected impact of thermal deformation on a VLBI delay observation for antenna ON and OE/OW following the model by \citet{Nothnagel_2009}.
This delay model uses mainly the antenna dimensions, the expansion coefficients of the material, and the temperature difference with respect to a reference temperature.
A temperature of 10~K higher than the reference temperature for the Onsala site is used for this graph.
A strong dependence on elevation is visible for both ON and OE/OW.
However, since the actual telescope towers have rather similar dimensions, the two curves are rather similar.
The largest differential effect for a delay observation on the ON-OE/OW baseline is on the order of 1.5~ps for an observation at zenith direction.
Table~\ref{TAB:investTD} presents the effect on the estimated weighted mean topocentric positions of OE and OW when including thermal deformation.
While the change is largest for the topocentric up-component, none of the changes are significant.

\begin{table}[hbt]
\centering
\caption{Effect of including thermal deformation modelling \citep{Nothnagel_2009} in the analysis. 
Presented are the corresponding changes in the weighted mean topocentric positions of OE and OW.}
\begin{tabular}{l | l  c c c }
antenna & $\Delta E$ (mm)     & $\Delta N$ (mm)     & $\Delta U$ (mm) \\
\hline
%OE & $-0.001 \pm 0.001$ & $+0.001 \pm 0.001$ & $+0.064 \pm 0.066$ \\ 
%OW & $-0.001 \pm 0.001$ & $+0.000 \pm 0.001$ & $+0.063 \pm 0.066$ \\ 
OE & $-0.001 \pm 0.002$ & $+0.001 \pm 0.001$ & $+0.038 \pm 0.079$ \\ 
OW & $-0.001 \pm 0.001$ & $+0.000 \pm 0.001$ & $+0.040 \pm 0.079$ \\ 
\hline
\end{tabular}
\label{TAB:investTD}
\end{table}

%- - - - - - - - - - - - - - - - - - - - - - - - - - - - - - - - -
% FIG-08
%- - - - - - - - - - - - - - - - - - - - - - - - - - - - - - - - -
%- - - - - - - - - - - - - - - - - - - - - - - - - - - - - - - - -

%- - - - - - - - - - - - - - - - - - - - - - - - - - - - - - - - - - - - 
\subsubsection{Impact of gravitational deformation}
\label{SEC:gravdef}
%- - - - - - - - - - - - - - - - - - - - - - - - - - - - - - - - - - - - 
Modelling of gravitational deformation of radio telescopes \citep{Nothnagel_et_al_2019, Loesler_et_al_2019} was not included in ITRF2014, but is strongly recommended by the IVS analysis coordinator, in particular for the preparations for analyses to prepare the upcoming ITRF2020 (John Gipson, private communication, 2020).
We therefore used {\em ASCOT} to investigate the impact of gravitational deformation on the analysis of the ONTIE-sessions.

\begin{figure*}[htb]
\centering
\includegraphics[width=1\textwidth]{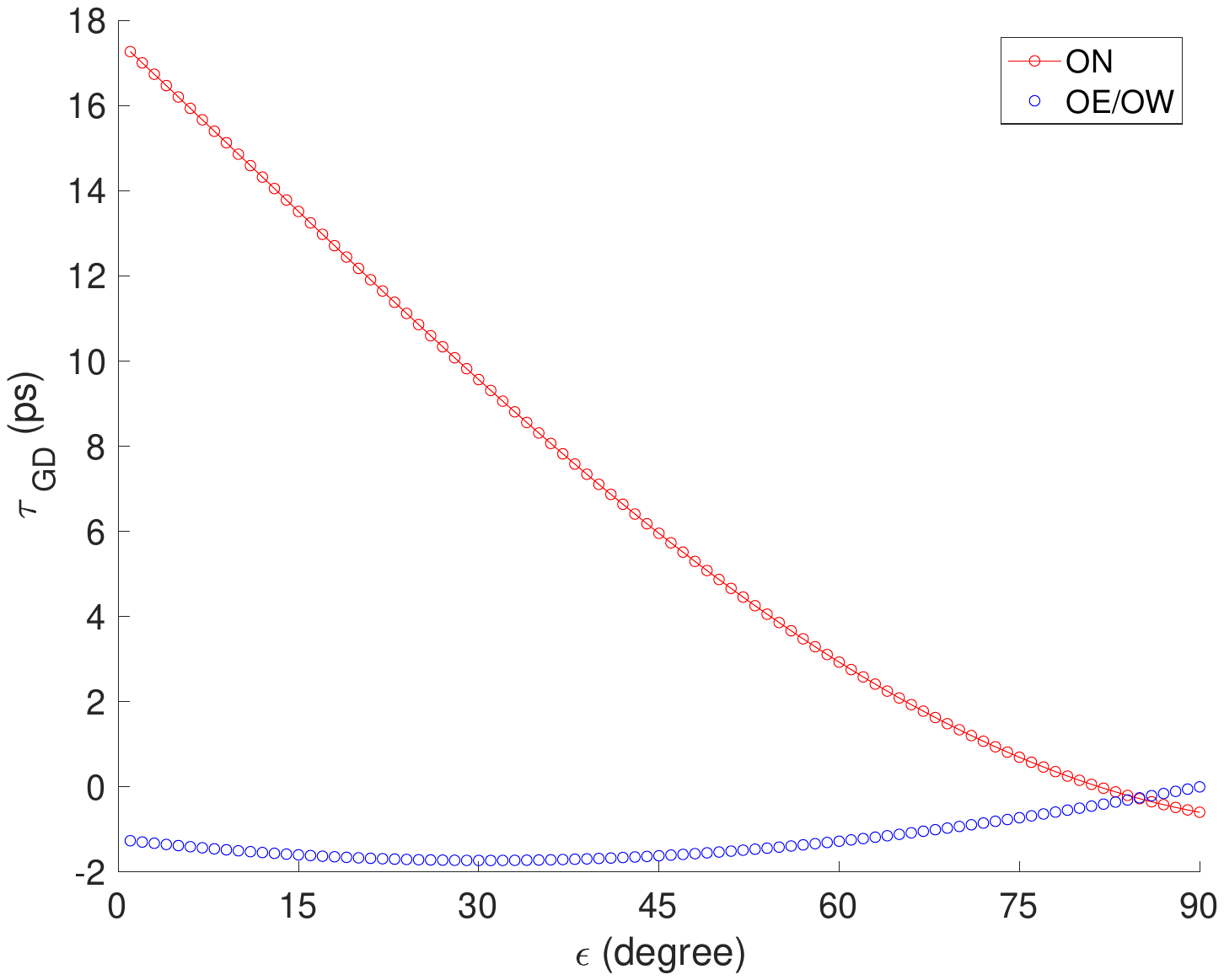}
\caption{Gravitational deformation effect on the VLBI delay   for the antennas ON \citep[red,][]{Nothnagel_et_al_2019} and OE/OW \citep[blue,][]{Loesler_et_al_2019}.}
\label{FIG:GD}      
\end{figure*}

Figure~\ref{FIG:GD} depicts the effect of gravitational deformation on the VLBI delay observation for the antennas ON \citep{Nothnagel_et_al_2019} and OE/OW \citep{Loesler_et_al_2019}.
Both antenna types again show a clear elevation dependence.
However, while OE/OW are rather stiff and stable antennas with a maximum effect of about 2~ps, ON is deforming much more and suffers from delay effects of almost 20~ps between observations in zenith and at the horizon.
The largest differential effect for a delay observation on the ON-OE/OW baseline is $\sim19$~ps for an observation at the horizon.
%- - - - - - - - - - - - - - - - - - - - - - - - - - - - - - - - -
% FIG-09
%- - - - - - - - - - - - - - - - - - - - - - - - - - - - - - - - -
%- - - - - - - - - - - - - - - - - - - - - - - - - - - - - - - - -

Table~\ref{TAB:investGD} presents the effects on the estimated weighted mean topocentric positions of OE and OW when these gravitational deformation effects are used or are not used.
Using or not using the gravitational deformation model for ON, see Tab.~\ref{TAB:investGD} a), changes the up-component significantly, more than 5.3~mm, while the horizontal changes are not significant. 
Using or not using the gravitational deformation model for OE and OW, see Tab.~\ref{TAB:investGD} b), also changes the up-component significantly, by more than 0.6~mm, while the horizontal changes are not significant.
%We note that modelling both thermal and gravitational deformation together changes the up-component of the positions determined for OE and OW by more than 5.4~mm. 
%This approach also gives the lowest WRMS scatter of the post-fit residuals.
These tests show the importance of modelling gravitational deformation of VLBI antennas in the data analysis.
We note that the consistent modelling of both thermal and gravitational deformation of all three antennas, ON, OE, OW, gives the lowest WRMS scatter of the post-fit residuals.

% - - - - - - - - - - - - - - - - - - - - - - - - - - - - 
\begin{table}[hbt]
\centering
\caption{Effect of including gravitational deformation modelling \citep{Nothnagel_et_al_2019, Loesler_et_al_2019} in the data analysis. 
Presented are the corresponding changes in the weighted mean topocentric positions of OE and OW when either a) the gravitational deformation model for ON is used or is not used, and b) when the gravitational deformation model for OE and OW is used or is not used.}
\begin{tabular}{l | l | c c c }
 & antenna & $\Delta E$ (mm)     & $\Delta N$ (mm)     & $\Delta U$ (mm) \\
\hline
a) & OE & $+0.003 \pm 0.002$ & $-0.014 \pm 0.003$ & $+5.387 \pm 0.009$ \\ 
   & OW & $+0.005 \pm 0.002$ & $-0.014 \pm 0.004$ & $+5.387 \pm 0.011$ \\
\hline
b) & OE & $+0.000 \pm 0.002$ & $-0.003 \pm 0.001$ & $+0.671 \pm 0.009$\\
   & OW & $-0.001 \pm 0.002$ & $-0.004 \pm 0.001$ & $+0.664 \pm 0.006$\\
\hline
\end{tabular}
\label{TAB:investGD}
\end{table}
%- - - - - - - - - - - - - - - - - - - - - - - - - - - - - 

%- - - - - - - - - - - - - - - - - - - - - - - - - - - - - - - - - - - - 
\section{Results}
\label{SEC:results}
%- - - - - - - - - - - - - - - - - - - - - - - - - - - - - - - - - - - - 
Our final (group-delay) positions for OE and OW are presented in Table~\ref{TAB:grpos-avg}. These were obtained as explained in Sect. \ref{SEC:globalGR}, as a global {\em ASCOT}  estimate combining all ONTIE sessions where phase- and cable calibration was present. 
% - - - - - - - - - - - - - - - - - - - - - - - - - - - - 
\begin{table}[htb]
\centering
\caption{VTRF2020b (epoch 2010.0) group-delay positions (in m) and their formal standard deviations (in mm) for OE and OW, obtained as described in Sect.~\ref{SEC:globalGR}.}
\begin{tabular}{ l | c | c | c | c | c | c}
antenna & X  & Y   & Z   & $\sigma_X$ & $\sigma_Y$ & $\sigma_Z$ \\
%antenna  & (m) & (m) & (m) & (mm) & (mm) & (mm) \\
\hline
OE & 3370889.29717 & 711571.19876 & 5349692.04692 & 0.11 & 0.05 & 0.17 \\
OW & 3370946.77840 & 711534.50648 & 5349660.92411 & 0.11 & 0.05 & 0.17 \\
\hline
\end{tabular}
\label{TAB:grpos-avg}
\end{table}
% - - - - - - - - - - - - - - - - - - - - - - - - - - - - 

% - - - - - - - - - - - - - - - - - - - - - - - - - - - - 
%\begin{table}[htb]
%\centering
%\caption{Local-tie vectors (in m) and their standard deviations %(in mm), obtained as described in Sect. \ref{SEC:oeowpos}.}
%\begin{tabular}{ l | c | c | c | c | c | c}
%Baseline & dX & dY & dZ & $\sigma_X$ & $\sigma_Y$ & $\sigma_Z$ %\\
%\hline
%OE-ON & 283.45679 & $-346.47713$ & $-138.82125$ & 0.32 & 0.09 & 0.41 \\
%OW-ON & 340.93799 & $-383.16943$ & $-169.94419$ & 0.26 & 0.09 & 0.37 \\
%\hline
%\end{tabular}
%\label{TAB:grvec-avg}
%\end{table}
% - - - - - - - - - - - - - - - - - - - - - - - - - - - - 

%- - - - - - - - - - - - - - - - - - - - - - - - - - - - - - - - - - - - 
\section{Summary and outlook}
\label{SEC:summary}
%- - - - - - - - - - - - - - - - - - - - - - - - - - - - - - - - - - - - 
During 2019 and 2020 we performed a series of short-baseline interferometry experiments to connect ONSA13NE and ONSA13SW, the Onsala twin telescopes, to the ONSALA60 VLBI station.
These co-location sessions were performed using X-band only observations.
The sessions were scheduled, observed, correlated and analysed at the Onsala Space Observatory. 
The coordinates of the Onsala twin telescopes could be determined in VTRF2020b with uncertainties on the sub-mm level.
This new set of coordinates (see Tab.~\ref{TAB:grpos-avg}) should be used from now on for scheduling, correlation, as a~priori for data analyses, and for comparison with classical local-tie techniques. We note, however, that since we use gravitational deformation modeling for all stations, the OTT positions derived will differ slightly with respect to the upcoming ITR2020 positions.
We note that the vgosDbs of all ONTIE- sessions are available via the IVS.

Using the {\em ASCOT} software, we investigated the impact of various effects such as thermal and gravitational deformation. We also compared phase-delay and group-delay position estimates, finding systematic offsets of about $\sim$3~mm in the Up-direction. This may be due to residual elevation-dependent delay-errors in the gravitational deformation model for ONSALA60, and/or other causes.

From the experience gained, we have developed procedures that allows us to repeat these kind of observations and the corresponding data processing and analysis on a regular basis.
Thus, regular monitoring of the baselines between the three stations can be done several times per year. 
These regular sessions can, with proper continuous amplitude calibration measurements, also provide regular flux-density monitoring observations of quasars.
This way of regularly checking the stability of the local network appears advantageous compared to performing labour-intensive classical local-tie surveys.

The plan is however to also perform a classical geodetic survey of the local tie vectors between the OTT and the reference points of the other geodetic stations at Onsala. This work was planned during early 2020, but was delayed due to the ongoing covid-19 pandemic.
Furthermore, we want to do a GNSS-based local-tie survey following the approach described in \citet{Ning_et_al_2015}.
We also aim at connecting the fourth VLBI station in the Onsala telescope cluster, the 25~m radio telescope, using similar interferometric measurements at C-band.

%- - - - - - - - - - - - - - - - - - - - - - - - - - - - - - - - -
\begin{acknowledgements}
%- - - - - - - - - - - - - - - - - - - - - - - - - - - - - - - - -
We want to acknowledge fruitful discussions and advise provided by Mike Titus (MIT/Haystack), Simone Bernhard (MPIfR Bonn), Grzegorz Klopotek (previously at Chalmers), and Sergei Bolotin (NVI).
%- - - - - - - - - - - - - - - - - - - - - - - - - - - - - - - - -
\end{acknowledgements}
%- - - - - - - - - - - - - - - - - - - - - - - - - - - - - - - - -

\section*{Declarations}
%=============================
\subsection*{Funding}
%=============================
Not applicable.

%------------------------------
\subsection*{Competing interests}
%------------------------------
The authors declare that there are no competing interests.

%-------------------------------------
\subsection*{Availability of data and material}
%-------------------------------------
The DiFX output and derived data products that support the findings of this study are available from the corresponding author upon reasonable request.
All vgosDbs are available via the IVS.

%------------------------------
\subsection*{Code availability}
%------------------------------
\ns{} is available via {https://sourceforge.net/projects/nusolve/}. \emph{ASCOT} is available via {https://github.com/varenius/ascot}.

%------------------------------
\subsection*{Authors' contributions}
%------------------------------
\noindent
EV and RH scheduled, observed, and correlated the observed data.
EV, RH, and TN analysed the data, using various software packages.
All authors contributed to writing the manuscript and read and approved the final version of the manuscript.

% BibTeX users please use one of
\bibliographystyle{spbasic}      % basic style, author-year citations
%\bibliographystyle{spmpsci}      % mathematics and physical sciences
%\bibliographystyle{spphys}       % APS-like style for physics
%\bibliography{}   % name your BibTeX data base
\bibliography{K-references}

\begin{thebibliography}{33}
\providecommand{\natexlab}[1]{#1}
\providecommand{\url}[1]{{#1}}
\providecommand{\urlprefix}{URL }
\expandafter\ifx\csname urlstyle\endcsname\relax
  \providecommand{\doi}[1]{DOI~\discretionary{}{}{}#1}\else
  \providecommand{\doi}{DOI~\discretionary{}{}{}\begingroup
  \urlstyle{rm}\Url}\fi
\providecommand{\eprint}[2][]{\url{#2}}

\bibitem[{Altamimi et~al.(2016)Altamimi, Rebischung, M\'etivier, and
  Collilieux}]{AltamimiITRF2014}
Altamimi Z, Rebischung P, M\'etivier L, Collilieux X (2016) {ITRF2014}: A new
  release of the {I}nternational {T}errestrial {R}eference {F}rame modeling
  nonlinear station motions. \emph{Journal of Geophysical Research: Solid
  Earth} 121:6109--6131, \doi{10.1002/2016JB013098}

\bibitem[{Artz et~al.(2016)Artz, Halsig, Iddink, and
  Nothnagel}]{Artz_et_al_2016}
Artz T, Halsig S, Iddink A, Nothnagel A (2016) ivg::ascot: Development of a new
  vlbi software package. In: Behrend D, Baver KD, Armstrong KL (eds) {\emph{IVS
  2016 General Meeting Proceedings "New Horizons with VGOS"}},
  {NASA/CP-2016-219016}, pp 217--221, \doi{10.22323/1.344.0140},
  \urlprefix\url{https://ivscc.gsfc.nasa.gov/publications/gm2016/045_artz_etal.pdf}

\bibitem[{Bachmann et~al.(2016)Bachmann, Thaller, Roggenbuck, L{\"o}sler, and
  Messerschmitt}]{Bachmann2016}
Bachmann S, Thaller D, Roggenbuck O, L{\"o}sler M, Messerschmitt L (2016) {{IVS
  contribution to ITRF2014}}. \emph{Journal of Geodesy} 90(7):631--654,
  \doi{10.1007/s00190-016-0899-4}

\bibitem[{Berg(1948)}]{Berg_1948}
Berg H (1948) Allgemeine Meteorologie; Einführung in die Physik der
  Atmosphäre. \emph{F.~Dümmler, Bonn}

\bibitem[{BKG(2020)}]{VTRF2020b}
BKG (2020) {VTRF2020b} combined solution.
  \urlprefix\url{https://ccivs.bkg.bund.de/combination/QUAT/SNX/VTRF2020b_IVS.snx},
  {Accessed: 2021-02-26}

\bibitem[{Bolotin et~al.(2012)Bolotin, Baver, Gipson, Gordon, and
  MacMillan}]{Bolotin_et_al_2012}
Bolotin S, Baver KD, Gipson J, Gordon D, MacMillan D (2012) The first release
  of $\nu$solve. In: Behrend D, Baver KD (eds) {\emph{IVS 2012 General Meeting
  Proceedings "Launching the Next-Generation IVS Network"}},
  {NASA/CP-2012-217504}, pp 222–--226,
  \urlprefix\url{http://ivscc.gsfc.nasa.gov/publications/gm2012/bolotin.pdf}

\bibitem[{Bolotin et~al.(2015)Bolotin, Baver, Gipson, Gordon, and
  MacMillan}]{Bolotin_et_al_2015}
Bolotin S, Baver KD, Gipson J, Gordon D, MacMillan D (2015) Implementation of
  the vgosdb format. In: Haas R, Colomer F (eds) {\emph{Proceedings of the 22nd
  European VLBI Group for Geodesy and Astrometry Working Meeting}}, pp
  150--152,
  \urlprefix\url{http://www.oso.chalmers.se/evga/22_EVGA_2015_Ponta_Delgada.pdf}

\bibitem[{{Charlot} et~al.(2020){Charlot}, {Jacobs}, {Gordon}, {Lambert}, {de
  Witt}, {B{\"o}hm}, {Fey}, {Heinkelmann}, {Skurikhina}, {Titov}, {Arias},
  {Bolotin}, {Bourda}, {Ma}, {Malkin}, {Nothnagel}, {Mayer}, {MacMillan},
  {Nilsson}, and {Gaume}}]{charlot2020}
{Charlot} P, {Jacobs} CS, {Gordon} D, {Lambert} S, {de Witt} A, {B{\"o}hm} J,
  {Fey} AL, {Heinkelmann} R, {Skurikhina} E, {Titov} O, {Arias} EF, {Bolotin}
  S, {Bourda} G, {Ma} C, {Malkin} Z, {Nothnagel} A, {Mayer} D, {MacMillan} DS,
  {Nilsson} T, {Gaume} R (2020) {The third realization of the International
  Celestial Reference Frame by very long baseline interferometry}. AAP
  644:A159, \doi{10.1051/0004-6361/202038368}, \eprint{2010.13625}

\bibitem[{Deller et~al.(2007)Deller, Tingay, Bailes, and West}]{Deller2007}
Deller AT, Tingay S, Bailes M, West C (2007) {DiFX}: {A} {S}oftware
  {C}orrelator for {V}ery {L}ong {B}aseline {I}nterferometry {U}sing
  {M}ulti-processor {C}omputing {E}nvironments. \emph{Publications of the
  Astronomical Society of the Pacific} 119(853):318--336, \doi{10.1086/513572}

\bibitem[{{Deller} et~al.(2011){Deller}, {Brisken}, {Phillips}, {Morgan},
  {Alef}, {Cappallo}, {Middelberg}, {Romney}, {Rottmann}, {Tingay}, and
  {Wayth}}]{Deller2011}
{Deller} AT, {Brisken} WF, {Phillips} CJ, {Morgan} J, {Alef} W, {Cappallo} R,
  {Middelberg} E, {Romney} J, {Rottmann} H, {Tingay} SJ, {Wayth} R (2011)
  {DiFX-2: A More Flexible, Efficient, Robust, and Powerful Software
  Correlator}. \emph{Publications of the Astronomical Society of the Pacific}
  123(901):275, \doi{10.1086/658907}

\bibitem[{Gipson(2007)}]{Gipson_2007}
Gipson J (2007) Correlation dur to station dependent noise in vlbi. In: Behrend
  D, Baver KD (eds) {\emph{IVS 2006 General Meeting Proceedings}},
  {NASA/CP-2006-214140}, pp 286--290,
  \urlprefix\url{https://ivscc.gsfc.nasa.gov/publications/gm2006/gipson.pdf}

\bibitem[{Gipson(2010)}]{Gipson2010}
Gipson J (2010) {An Introduction to Sked}. In: Behrend D, Baver KD (eds)
  {\emph{IVS 2010 General Meeting Proceedings}}, {International VLBI Service
  for Geodesy and Astrometry}, pp 77--84

\bibitem[{Haas(2013)}]{Haas_2013}
Haas R (2013) {The Onsala twin telescope project}. In: Zubko N, Poutanen M
  (eds) {\emph{Proceedings of the 21st European VLBI for Geodesy and Astrometry
  (EVGA) working meeting}}, {ISBN: 978-9517112963}, pp 61--66

\bibitem[{Haas and Eschelbach(2005)}]{Haas_Eschelbach_2005}
Haas R, Eschelbach C (2005) {The 2002 Local Tie Survey at the Onsala Space
  Observatory}. IERS Technical Note No 33 pp 55--63

\bibitem[{Haas et~al.(2019)Haas, Casey, Conway, Elgered, Hammargren, Helldner,
  Johansson, Kylenfall, Lerner, Pettersson, and
  Wennerb{\"a}ck}]{Haas_et_al_2019}
Haas R, Casey S, Conway J, Elgered G, Hammargren R, Helldner L, Johansson
  K{\AA}, Kylenfall U, Lerner M, Pettersson L, Wennerb{\"a}ck L (2019) {Status
  of the Onsala Twin Telescopes — Two Years After the Inauguration}. In: Haas
  R, Garcia-Espada S, Lopez~Fernandez JA (eds) {\emph{Proceedings of the 24th
  European VLBI for Geodesy and Astrometry (EVGA) working meeting}}, {ISBN:
  978-84-416-5634-5}, pp 5--9

\bibitem[{L{\"o}sler et~al.(2013)L{\"o}sler, Haas, and
  Eschelbach}]{Loesler_et_al_2013}
L{\"o}sler M, Haas R, Eschelbach C (2013) {{Automated and continual
  determination of radio telescope reference points with sub-mm accuracy:
  results from a campaign at the Onsala Space Observatory}}. \emph{Journal of
  Geodesy} 87:791–804, \doi{10.1007/s00190-013-0647-y}

\bibitem[{L{\"o}sler et~al.(2016)L{\"o}sler, Haas, and
  Eschelbach}]{Loesler_et_al_2016}
L{\"o}sler M, Haas R, Eschelbach C (2016) {{Terrestrial monitoring of a radio
  telescope reference point using comprehensive uncertainty budgeting --
  Investigations during CONT14 at the Onsala Space Observatory}}. \emph{Journal
  of Geodesy} 90:467–486, \doi{10.1007/s00190-016-0887-8}

\bibitem[{L{\"o}sler et~al.(2019)L{\"o}sler, Eschelbach, Haas, and
  Greiwe}]{Loesler_et_al_2019}
L{\"o}sler M, Eschelbach C, Haas R, Greiwe A (2019) {{Gravitational deformation
  of ring-focus antennas for VGOS: first investigations at the Onsala twin
  telescopes project}}. \emph{Journal of Geodesy} 93:2069–2087,
  \doi{10.1007/s00190-019-01302-5}

\bibitem[{Markn{\"a}s(2019)}]{Marknas_2019}
Markn{\"a}s V (2019) {Connecting the Onsala Telescope Cluster Using Local
  Interferometry}. Master's thesis, Chalmers University of Technology

\bibitem[{MIT/Haystack(2020)}]{HOPSweb}
MIT/Haystack (2020) Haystack observatory postprocessing system (hops).
  \urlprefix\url{https://www.haystack.mit.edu/haystack-observatory-postprocessing-system-hops/},
  {Accessed: 2020-08-27}

\bibitem[{Ning et~al.(2015)Ning, Haas, and Elgered}]{Ning_et_al_2015}
Ning T, Haas R, Elgered G (2015) {Determination of the local tie vector between
  the VLBI and GNSS reference points at Onsala using GPS measurements}.
  \emph{Journal of Geodesy} 89:711–723, \doi{10.1007/s00190-015-0809-1}

\bibitem[{Nothnagel(2009)}]{Nothnagel_2009}
Nothnagel A (2009) {Conventions on thermal expansion modelling of radio
  telescopes for geodetic and astrometric VLBI}. \emph{Journal of Geodesy}
  83:787–792, \doi{10.1007/s00190-008-0284-z}

\bibitem[{Nothnagel et~al.(2019)Nothnagel, Holst, and
  Haas}]{Nothnagel_et_al_2019}
Nothnagel A, Holst C, Haas R (2019) {A VLBI delay model for gravitational
  deformations of the Onsala 20 m radio telescope and the impact on its global
  coordinates}. \emph{Journal of Geodesy} 93:2019–2036,
  \doi{10.1007/s00190-019-01299-x}

\bibitem[{Petrachenko et~al.(2009)Petrachenko, Niell, Behrend, Corey, B{\"o}hm,
  Charlot, Collioud, Gipson, Haas, Hobiger, Koyama, MacMillan, Malkin, Nilsson,
  Pany, Tuccari, Whitney, and Wresnik}]{Petrachenko_et_al_2009}
Petrachenko B, Niell A, Behrend D, Corey B, B{\"o}hm J, Charlot P, Collioud A,
  Gipson J, Haas R, Hobiger T, Koyama Y, MacMillan D, Malkin Z, Nilsson T, Pany
  A, Tuccari G, Whitney A, Wresnik J (2009) {Design aspects of the VLBI2010
  system}. Progress report of the IVS VLBI2010 committee

\bibitem[{Plag and Pearlman(2009)}]{Plag2009}
Plag HP, Pearlman M (eds)  (2009) Global Geodetic Observing System: Meeting the
  Requirements of a Global Society on a Changing Planet in 2020, 1st edn.
  \emph{Springer-Verlag Berlin Heidelberg}, \doi{10.1007/978-3-642-02687-4}

\bibitem[{re3data.org(2021)}]{VMF3}
re3dataorg (2021) Vmf data server; editing status 2020-12-14.
  \doi{10.17616/R3RD2H}, \urlprefix\url{http://doi.org/10.17616/R3RD2H},
  {Accessed: 2021-02-26}

\bibitem[{Scherneck et~al.(1998)Scherneck, Elgered, Johansson, and
  R{\"o}nn{\"a}ng}]{Scherneck_et_al_1998}
Scherneck HG, Elgered G, Johansson JM, R{\"o}nn{\"a}ng BO (1998) {Space
  Geodetic Activities at the Onsala Space Observatory: 25 years in the Service
  of Plate Tectonics}. \emph{Physics and Chemistry of the Earth} {23
  }({7-8}):811–823, \doi{10.1016/S0079-1946(98)00094-9}

\bibitem[{Thaller(2021)}]{Thaller_2021}
Thaller D (2021) personal communication, 2021-02-19

\bibitem[{Tuccari et~al.(2010)Tuccari, Alef, Bertarini, Buttaccio, Comoretto,
  Graham, Neidhardt, Platania, Russo, Roy, Wunderlich, Zeitlh{\"o}fler, and
  Xiang}]{Tuccari_DBBC2_2010}
Tuccari G, Alef W, Bertarini A, Buttaccio S, Comoretto G, Graham D, Neidhardt
  A, Platania PR, Russo A, Roy A, Wunderlich M, Zeitlh{\"o}fler R, Xiang Y
  (2010) {DBBC2 Backend: Status and Development Plan}. In: Behrend D, Baver KD
  (eds) {\emph{IVS 2010 General Meeting Proceedings “VLBI2010: From Vision to
  Reality”}}, {NASA/CP-2010-215864}, pp 392--395, \doi{10.22323/1.344.0140},
  \urlprefix\url{http://ivscc.gsfc.nasa.gov/publications/gm2010/tuccari2.pdf}

\bibitem[{Tuccari et~al.(2018)Tuccari, Alef, Dornbusch, Wunderlich, Roy,
  Wagner, Haas, and Johansson}]{Tuccari_DBBC3_2018}
Tuccari G, Alef W, Dornbusch S, Wunderlich M, Roy A, Wagner J, Haas R,
  Johansson K{\AA} (2018) {DBBC3 — the new wide-band backend for VLBI}. In:
  {\emph{Proceedings of the 14th European VLBI Network Symposium \& Users
  Meeting (EVN 2018) 8-11 October 2018}}, {PoS - Proceedings of Science},
  \doi{10.22323/1.344.0140}, \urlprefix\url{https://pos.sissa.it/344/140/pdf}

\bibitem[{UN(2017)}]{UN2017}
UN (2017) A global geodetic reference frame for sustainable development.
  \urlprefix\url{https://www.un.org/ga/search/view_doc.asp?symbol=A/69/L.53}

\bibitem[{Verkouter(2020)}]{jive5abweb}
Verkouter H (2020) jive5ab on github.
  \urlprefix\url{https://github.com/jive-vlbi/jive5ab}, {Accessed: 2020-08-27}

\bibitem[{Whitney(1974)}]{Whitney_1974}
Whitney AR (1974) {Precision Geodesy and Astrometry via Very Long Baseline
  Interferometry}. PhD thesis, MIT Cambridge

\end{thebibliography}

\appendix

%------------------------------------------------
\section{Positions of OE and OW for all vgosDbs}
\label{SEC:posapp}
%------------------------------------------------
This appendix contains OE and OW positions obtained with {\emph ASCOT} for all vgosDbs. Group-delay results are presented in Tables \ref{TAB:gr-oepos} and \ref{TAB:gr-owpos}, and phase-delay results in Tables \ref{TAB:ph-oepos} and \ref{TAB:ph-owpos}. These data can be used, together with Table \ref{TAB:grpos-avg}, to reproduce Figures \ref{fig:GR_vs_time} and \ref{fig:PH_vs_time}. The uncertainties are the formal errors given by {\emph ASCOT}.

\begin{landscape}
\begin{table}[htb]
\centering
\caption{VTRF2020b (epoch 2010.0) group-delay positions for OE. The reference date and time for the estimated position is also listed, as well as the formal uncertainties reported by {\em ASCOT}.}
\begin{tabular}{ c | c | c | c | c | c | c | c | c | c}
VGOSDB &  Ref. time (ISO 8601)  & X (m)& Y (m)& Z (m)& $\sigma_X$(mm) & $\sigma_Y$ (mm)& $\sigma_Z$ (mm) &  WRMS (ps)\\
% &  &  & [m]& [m] & [m] & [mm]& [mm]& [mm]\\
\hline
19APR24VB & 2019-04-24T19:45:29 & 3370889.30129 & 711571.19968 & 5349692.06 & 2.03 & 0.98 & 3.59 & 11.58 \\
19APR30VB & 2019-05-01T10:12:30 & 3370889.30016 & 711571.19984 & 5349692.05 & 0.44 & 0.19 & 0.67 & 10.21 \\
19MAY02VB & 2019-05-02T15:42:29 & 3370889.29924 & 711571.19917 & 5349692.05 & 0.48 & 0.22 & 0.71 & 9.21 \\
19MAY15VB & 2019-05-16T11:35:00 & 3370889.30062 & 711571.19920 & 5349692.05 & 1.14 & 0.56 & 1.78 & 12.25 \\
19MAY16VB & 2019-05-17T02:59:29 & 3370889.29849 & 711571.19987 & 5349692.05 & 0.55 & 0.26 & 0.83 & 14.15 \\
19MAY22VB & 2019-05-22T20:59:29 & 3370889.30042 & 711571.20017 & 5349692.05 & 1.08 & 0.45 & 1.68 & 11.80 \\
19NOV19VB & 2019-11-20T04:29:29 & 3370889.29785 & 711571.19895 & 5349692.05 & 0.52 & 0.27 & 0.79 & 12.07 \\
19NOV23VB & 2019-11-23T19:07:00 & 3370889.29949 & 711571.19896 & 5349692.05 & 0.43 & 0.21 & 0.65 & 10.60 \\
19NOV24VB & 2019-11-24T19:07:30 & 3370889.29999 & 711571.19916 & 5349692.05 & 0.47 & 0.21 & 0.71 & 11.28 \\
20JAN10VB & 2020-01-11T04:36:00 & 3370889.29638 & 711571.19918 & 5349692.05 & 0.59 & 0.29 & 0.94 & 13.89 \\
20JAN11VB & 2020-01-12T00:49:30 & 3370889.29633 & 711571.19859 & 5349692.05 & 0.68 & 0.32 & 1.03 & 16.18 \\
20JAN12VB & 2020-01-12T20:42:59 & 3370889.29686 & 711571.19902 & 5349692.05 & 0.62 & 0.31 & 0.95 & 14.73 \\
20MAR19VB & 2020-03-20T03:59:30 & 3370889.29502 & 711571.19870 & 5349692.04 & 0.40 & 0.19 & 0.62 & 9.62 \\
20MAR20VB & 2020-03-21T04:29:00 & 3370889.29373 & 711571.19841 & 5349692.04 & 0.39 & 0.19 & 0.60 & 9.77 \\
20MAR21VB & 2020-03-22T04:59:30 & 3370889.29526 & 711571.19877 & 5349692.04 & 0.40 & 0.20 & 0.61 & 10.01 \\
20MAR22VB & 2020-03-23T04:59:30 & 3370889.29639 & 711571.19830 & 5349692.05 & 0.41 & 0.19 & 0.62 & 9.91 \\
20JUN25VB & 2020-06-26T05:37:30 & 3370889.29807 & 711571.19906 & 5349692.05 & 0.43 & 0.22 & 0.66 & 10.21 \\
20JUN26VB & 2020-06-27T03:52:30 & 3370889.29881 & 711571.19924 & 5349692.05 & 0.44 & 0.21 & 0.65 & 10.41 \\
20JUN27VB & 2020-06-28T02:06:30 & 3370889.29866 & 711571.19906 & 5349692.05 & 0.48 & 0.21 & 0.70 & 10.98 \\
20JUN28VB & 2020-06-29T01:05:59 & 3370889.29652 & 711571.19863 & 5349692.05 & 0.50 & 0.24 & 0.75 & 12.21 \\
20AUG10VB & 2020-08-11T00:59:30 & 3370889.29780 & 711571.19890 & 5349692.05 & 0.54 & 0.26 & 0.82 & 13.84 \\
20AUG14VB & 2020-08-15T06:06:59 & 3370889.29739 & 711571.19903 & 5349692.05 & 0.49 & 0.25 & 0.75 & 12.39 \\
20AUG15VB & 2020-08-16T06:06:59 & 3370889.29834 & 711571.19854 & 5349692.05 & 0.51 & 0.24 & 0.78 & 12.19 \\
20NOV12VB & 2020-11-12T18:59:30 & 3370889.29797 & 711571.19773 & 5349692.05 & 0.61 & 0.29 & 0.94 & 15.05 \\
20NOV13VB & 2020-11-13T18:59:30 & 3370889.29737 & 711571.19919 & 5349692.05 & 0.61 & 0.29 & 0.93 & 17.14 \\
\end{tabular}
\label{TAB:gr-oepos}
\end{table}
\end{landscape}

\begin{landscape}
\begin{table}[htb]
\centering
\caption{VTRF2020b (epoch 2010.0) group-delay positions for OW. The reference date and time for the estimated position is also listed, as well as the formal uncertainties reported by {\em ASCOT}.}
\begin{tabular}{ c | c | c | c | c | c | c | c | c | c}
VGOSDB &  Ref. time (ISO 8601)  & X (m)& Y (m)& Z (m)& $\sigma_X$(mm) & $\sigma_Y$ (mm)& $\sigma_Z$ (mm) &  WRMS (ps)\\
% &  &  & [m]& [m] & [m] & [mm]& [mm]& [mm]\\
\hline
19APR30VB & 2019-05-01T10:12:30 & 3370946.78280 & 711534.50749 & 5349660.93 & 0.44 & 0.19 & 0.67 & 10.21 \\
19MAY15VB & 2019-05-16T11:35:00 & 3370946.77909 & 711534.50801 & 5349660.93 & 1.19 & 0.58 & 1.85 & 12.25 \\
19MAY16VB & 2019-05-17T02:59:29 & 3370946.77803 & 711534.50684 & 5349660.92 & 0.60 & 0.28 & 0.90 & 14.15 \\
19MAY22VB & 2019-05-22T20:59:29 & 3370946.78143 & 711534.50726 & 5349660.93 & 1.12 & 0.47 & 1.75 & 11.80 \\
19NOV19VB & 2019-11-20T04:29:29 & 3370946.78039 & 711534.50731 & 5349660.93 & 0.52 & 0.27 & 0.80 & 12.07 \\
19NOV23VB & 2019-11-23T19:07:00 & 3370946.78026 & 711534.50712 & 5349660.93 & 0.43 & 0.21 & 0.65 & 10.60 \\
19NOV24VB & 2019-11-24T19:07:30 & 3370946.78083 & 711534.50734 & 5349660.93 & 0.47 & 0.21 & 0.71 & 11.28 \\
20JAN10VB & 2020-01-11T04:36:00 & 3370946.77851 & 711534.50688 & 5349660.92 & 0.60 & 0.29 & 0.95 & 13.89 \\
20JAN11VB & 2020-01-12T00:49:30 & 3370946.77845 & 711534.50618 & 5349660.92 & 0.69 & 0.33 & 1.05 & 16.18 \\
20JAN12VB & 2020-01-12T20:42:59 & 3370946.77781 & 711534.50617 & 5349660.92 & 0.64 & 0.31 & 0.96 & 14.73 \\
20MAR19VB & 2020-03-20T03:59:30 & 3370946.77769 & 711534.50599 & 5349660.92 & 0.41 & 0.19 & 0.63 & 9.62 \\
20MAR20VB & 2020-03-21T04:29:00 & 3370946.77762 & 711534.50611 & 5349660.92 & 0.40 & 0.19 & 0.61 & 9.77 \\
20MAR21VB & 2020-03-22T04:59:30 & 3370946.77736 & 711534.50642 & 5349660.92 & 0.41 & 0.20 & 0.63 & 10.01 \\
20MAR22VB & 2020-03-23T04:59:30 & 3370946.77902 & 711534.50600 & 5349660.92 & 0.42 & 0.20 & 0.64 & 9.91 \\
20JUN25VB & 2020-06-26T05:37:30 & 3370946.77810 & 711534.50605 & 5349660.92 & 0.43 & 0.22 & 0.67 & 10.21 \\
20JUN26VB & 2020-06-27T03:52:30 & 3370946.77881 & 711534.50637 & 5349660.92 & 0.45 & 0.21 & 0.66 & 10.41 \\
20JUN27VB & 2020-06-28T02:06:30 & 3370946.77928 & 711534.50694 & 5349660.93 & 0.49 & 0.21 & 0.70 & 10.98 \\
20JUN28VB & 2020-06-29T01:05:59 & 3370946.77705 & 711534.50609 & 5349660.92 & 0.51 & 0.24 & 0.76 & 12.21 \\
20AUG10VB & 2020-08-11T00:59:30 & 3370946.77794 & 711534.50674 & 5349660.92 & 0.58 & 0.28 & 0.87 & 13.84 \\
20AUG14VB & 2020-08-15T06:06:59 & 3370946.77754 & 711534.50647 & 5349660.92 & 0.50 & 0.25 & 0.76 & 12.39 \\
20AUG15VB & 2020-08-16T06:06:59 & 3370946.77810 & 711534.50643 & 5349660.92 & 0.52 & 0.25 & 0.79 & 12.19 \\
20NOV12VB & 2020-11-12T18:59:30 & 3370946.77849 & 711534.50644 & 5349660.93 & 0.62 & 0.30 & 0.97 & 15.05 \\
20NOV13VB & 2020-11-13T18:59:30 & 3370946.77740 & 711534.50654 & 5349660.92 & 0.62 & 0.30 & 0.95 & 17.14 \\
\end{tabular}
\label{TAB:gr-owpos}
\end{table}
\end{landscape}

\begin{landscape}
\begin{table}[htb]
\centering
\caption{VTRF2020b (epoch 2010.0) phase-delay positions for OE. The reference date and time for the estimated position is also listed, as well as the formal uncertainties reported by {\em ASCOT}.}
\begin{tabular}{ c | c | c | c | c | c | c | c | c | c}
VGOSDB &  Ref. time (ISO 8601)  & X (m)& Y (m)& Z (m)& $\sigma_X$(mm) & $\sigma_Y$ (mm)& $\sigma_Z$ (mm) &  WRMS (ps)\\
% &  &  & [m]& [m] & [m] & [mm]& [mm]& [mm]\\
\hline
19APR24VB & 2019-04-24T19:45:29 & 3370889.30778 & 711571.20103 & 5349692.07 & 1.71 & 0.74 & 2.81 & 9.01 \\
19APR30VB & 2019-05-01T10:12:30 & 3370889.30532 & 711571.20019 & 5349692.06 & 0.35 & 0.16 & 0.54 & 7.42 \\
19MAY02VB & 2019-05-02T15:42:29 & 3370889.30400 & 711571.20024 & 5349692.06 & 0.58 & 0.26 & 0.88 & 10.08 \\
19MAY15VB & 2019-05-16T11:35:00 & 3370889.30311 & 711571.20176 & 5349692.06 & 0.71 & 0.30 & 1.07 & 5.97 \\
19MAY16VB & 2019-05-17T02:59:29 & 3370889.30287 & 711571.20009 & 5349692.06 & 0.32 & 0.15 & 0.50 & 6.96 \\
19MAY22VB & 2019-05-22T20:59:29 & 3370889.30810 & 711571.19977 & 5349692.06 & 0.84 & 0.35 & 1.32 & 7.63 \\
19NOV19VB & 2019-11-20T04:29:29 & 3370889.29981 & 711571.19910 & 5349692.05 & 0.13 & 0.06 & 0.20 & 2.69 \\
19NOV23VB & 2019-11-23T19:07:00 & 3370889.30021 & 711571.19911 & 5349692.05 & 0.11 & 0.05 & 0.17 & 2.48 \\
19NOV24VB & 2019-11-24T19:07:30 & 3370889.30033 & 711571.19924 & 5349692.05 & 0.12 & 0.05 & 0.19 & 2.59 \\
20JAN10VB & 2020-01-11T04:36:00 & 3370889.29939 & 711571.19915 & 5349692.05 & 0.16 & 0.08 & 0.26 & 3.33 \\
20JAN11VB & 2020-01-12T00:49:30 & 3370889.29896 & 711571.19895 & 5349692.05 & 0.15 & 0.06 & 0.23 & 3.01 \\
20JAN12VB & 2020-01-12T20:42:59 & 3370889.29967 & 711571.19916 & 5349692.05 & 0.15 & 0.07 & 0.23 & 2.98 \\
20MAR19VB & 2020-03-20T03:59:30 & 3370889.29710 & 711571.19907 & 5349692.05 & 0.12 & 0.06 & 0.19 & 2.73 \\
20MAR20VB & 2020-03-21T04:29:00 & 3370889.29702 & 711571.19878 & 5349692.05 & 0.13 & 0.06 & 0.19 & 2.84 \\
20MAR21VB & 2020-03-22T04:59:30 & 3370889.29692 & 711571.19890 & 5349692.05 & 0.13 & 0.06 & 0.20 & 2.93 \\
20MAR22VB & 2020-03-23T04:59:30 & 3370889.29812 & 711571.19894 & 5349692.05 & 0.15 & 0.07 & 0.23 & 3.16 \\
20JUN25VB & 2020-06-26T05:37:30 & 3370889.30058 & 711571.19977 & 5349692.05 & 0.19 & 0.09 & 0.29 & 3.85 \\
20JUN26VB & 2020-06-27T03:52:30 & 3370889.29984 & 711571.19926 & 5349692.05 & 0.20 & 0.09 & 0.29 & 3.96 \\
20JUN27VB & 2020-06-28T02:06:30 & 3370889.29978 & 711571.19947 & 5349692.05 & 0.23 & 0.09 & 0.33 & 4.54 \\
20JUN28VB & 2020-06-29T01:05:59 & 3370889.29900 & 711571.19932 & 5349692.05 & 0.28 & 0.12 & 0.42 & 5.85 \\
20AUG10VB & 2020-08-11T00:59:30 & 3370889.30260 & 711571.20095 & 5349692.06 & 0.33 & 0.15 & 0.50 & 6.98 \\
20AUG14VB & 2020-08-15T06:06:59 & 3370889.29922 & 711571.19967 & 5349692.05 & 0.16 & 0.07 & 0.25 & 3.43 \\
20AUG15VB & 2020-08-16T06:06:59 & 3370889.29994 & 711571.19934 & 5349692.05 & 0.15 & 0.07 & 0.24 & 3.19 \\
20NOV12VB & 2020-11-12T18:59:30 & 3370889.30098 & 711571.19990 & 5349692.05 & 0.13 & 0.06 & 0.20 & 2.87 \\
20NOV13VB & 2020-11-13T18:59:30 & 3370889.30117 & 711571.20008 & 5349692.05 & 0.15 & 0.07 & 0.24 & 3.82 \\
\end{tabular}
\label{TAB:ph-oepos}
\end{table}
\end{landscape}

\begin{landscape}
\begin{table}[htb]
\centering
\caption{VTRF2020b (epoch 2010.0) phase-delay positions for OW. The reference date and time for the estimated position is also listed, as well as the formal uncertainties reported by {\em ASCOT}.}
\begin{tabular}{ c | c | c | c | c | c | c | c | c | c}
VGOSDB &  Ref. time (ISO 8601)  & X (m)& Y (m)& Z (m)& $\sigma_X$(mm) & $\sigma_Y$ (mm)& $\sigma_Z$ (mm) &  WRMS (ps)\\
% &  &  & [m]& [m] & [m] & [mm]& [mm]& [mm]\\
\hline
19APR30VB & 2019-05-01T10:12:30 & 3370946.78620 & 711534.50762 & 5349660.94 & 0.35 & 0.16 & 0.54 & 7.42 \\
19MAY15VB & 2019-05-16T11:35:00 & 3370946.78042 & 711534.50906 & 5349660.93 & 0.71 & 0.30 & 1.08 & 5.97 \\
19MAY16VB & 2019-05-17T02:59:29 & 3370946.78081 & 711534.50722 & 5349660.93 & 0.32 & 0.15 & 0.50 & 6.96 \\
19MAY22VB & 2019-05-22T20:59:29 & 3370946.78852 & 711534.50730 & 5349660.94 & 0.84 & 0.35 & 1.32 & 7.63 \\
19NOV19VB & 2019-11-20T04:29:29 & 3370946.78026 & 711534.50664 & 5349660.93 & 0.13 & 0.06 & 0.20 & 2.69 \\
19NOV23VB & 2019-11-23T19:07:00 & 3370946.78090 & 711534.50684 & 5349660.93 & 0.11 & 0.05 & 0.17 & 2.48 \\
19NOV24VB & 2019-11-24T19:07:30 & 3370946.78087 & 711534.50682 & 5349660.93 & 0.12 & 0.05 & 0.19 & 2.59 \\
20JAN10VB & 2020-01-11T04:36:00 & 3370946.77988 & 711534.50670 & 5349660.93 & 0.16 & 0.08 & 0.26 & 3.33 \\
20JAN11VB & 2020-01-12T00:49:30 & 3370946.77963 & 711534.50648 & 5349660.93 & 0.15 & 0.06 & 0.23 & 3.01 \\
20JAN12VB & 2020-01-12T20:42:59 & 3370946.78036 & 711534.50673 & 5349660.93 & 0.15 & 0.07 & 0.23 & 2.98 \\
20MAR19VB & 2020-03-20T03:59:30 & 3370946.78020 & 711534.50665 & 5349660.93 & 0.12 & 0.06 & 0.19 & 2.73 \\
20MAR20VB & 2020-03-21T04:29:00 & 3370946.77956 & 711534.50635 & 5349660.93 & 0.13 & 0.06 & 0.19 & 2.84 \\
20MAR21VB & 2020-03-22T04:59:30 & 3370946.77940 & 711534.50638 & 5349660.93 & 0.13 & 0.06 & 0.20 & 2.93 \\
20MAR22VB & 2020-03-23T04:59:30 & 3370946.78075 & 711534.50653 & 5349660.93 & 0.15 & 0.07 & 0.23 & 3.16 \\
20JUN25VB & 2020-06-26T05:37:30 & 3370946.77946 & 711534.50655 & 5349660.93 & 0.19 & 0.09 & 0.29 & 3.85 \\
20JUN26VB & 2020-06-27T03:52:30 & 3370946.77886 & 711534.50612 & 5349660.93 & 0.20 & 0.09 & 0.29 & 3.96 \\
20JUN27VB & 2020-06-28T02:06:30 & 3370946.77888 & 711534.50638 & 5349660.92 & 0.23 & 0.09 & 0.33 & 4.54 \\
20JUN28VB & 2020-06-29T01:05:59 & 3370946.77849 & 711534.50635 & 5349660.92 & 0.28 & 0.12 & 0.41 & 5.85 \\
20AUG10VB & 2020-08-11T00:59:30 & 3370946.78146 & 711534.50769 & 5349660.93 & 0.33 & 0.15 & 0.50 & 6.98 \\
20AUG14VB & 2020-08-15T06:06:59 & 3370946.77812 & 711534.50652 & 5349660.92 & 0.16 & 0.08 & 0.25 & 3.43 \\
20AUG15VB & 2020-08-16T06:06:59 & 3370946.77879 & 711534.50621 & 5349660.93 & 0.16 & 0.07 & 0.24 & 3.19 \\
20NOV12VB & 2020-11-12T18:59:30 & 3370946.78053 & 711534.50688 & 5349660.93 & 0.13 & 0.06 & 0.20 & 2.87 \\
20NOV13VB & 2020-11-13T18:59:30 & 3370946.78061 & 711534.50710 & 5349660.93 & 0.15 & 0.07 & 0.24 & 3.82 \\
\end{tabular}
\label{TAB:ph-owpos}
\end{table}
\end{landscape}
% - - - - - - - - - - - - - - - - - - - - - - - - - - - - 
%%%%%%%%%%%%%%%%%%%%%%%%%%%%%%%%%%%%%%%%%%%%%%%%%%%%%%%%%%%%%%%%%%%%%%%%%%%%%
\end{document}